\documentclass[5p,letter]{elsarticle}
\usepackage[utf8]{inputenc}

\usepackage{amsmath}
\usepackage{amssymb}
\usepackage{booktabs}
\usepackage{collcell}
\usepackage[T1]{fontenc}
\usepackage{graphicx}    
\usepackage[version=4]{mhchem}
\usepackage{setspace}
\usepackage{subcaption}
\usepackage[table]{xcolor}

\makeatletter
\def\ps@pprintTitle{%
 \let\@oddhead\@empty
 \let\@evenhead\@empty
 \def\@oddfoot{}%
 \let\@evenfoot\@oddfoot}
\makeatother

\usepackage{siunitx}
\usepackage{geometry}
\usepackage[textwidth=1.0in]{todonotes} 
\setuptodonotes{size=\scriptsize}
\makeatletter
\renewcommand{\todo}[2][]{%
    \@todo[caption={#2}, #1]{\begin{spacing}{0.5}#2\end{spacing}}%
} 
\makeatother

\usepackage[colorlinks=true,citecolor=blue,urlcolor=blue]{hyperref}
\usepackage[capitalise]{cleveref}

\newcommand\AddLabel[1]{%
  \refstepcounter{equation}% increment equation counter
  (\theequation)% print equation number
  \label{#1}% give the equation a \label
}
\newcolumntype{M}{>{$\displaystyle}l<{$}} % mathematics column
\newcolumntype{L}{>{\collectcell\AddLabel}r<{\endcollectcell}}

\journal{Journal of Power Sources}

\begin{document}

\begin{frontmatter}

\title{A Pseudo-Two-Dimensional (P2D) Model for \ce{FeS2} Conversion Cathode Batteries} 

\author[1]{Jeffrey S. Horner}
\author[2]{Grace Whang}
\author[3]{Igor V. Kolesnichenko}
\author[3]{Timothy N. Lambert}
\author[2]{Bruce S. Dunn}
\author[1]{Scott A. Roberts\corref{cor1}}
\ead{sarober@sandia.gov}
\cortext[cor1]{Corresponding author}
\address[1]{Thermal/Fluid Component Sciences Department, Sandia National Laboratories, Albuquerque, New Mexico, USA}
\address[2]{Materials Science and Engineering Department, University of California, Los Angeles, Los Angeles, California, USA}
\address[3]{Photovoltaics and Materials Technology Department, Sandia National Laboratories, Albuquerque, New Mexico, USA}

\begin{abstract}

Conversion cathode materials are gaining interest for secondary batteries due to their high theoretical energy and power density. However, practical application as a secondary battery material is currently limited by practical issues such as poor cyclability. To better understand these materials, we have developed a pseudo-two-dimensional model for conversion cathodes. We apply this model to \ce{FeS2} -- a material that undergoes intercalation followed by conversion during discharge. The model is derived from the half-cell Doyle-Fuller-Newman model with additional loss terms added to reflect the converted shell resistance as the reaction progresses. We also account for polydisperse active material particles by incorporating a variable active surface area and effective particle radius. Using the model, we show that the leading loss mechanisms for \ce{FeS2} are associated with solid-state diffusion and electrical transport limitations through the converted shell material. The polydisperse simulations are also compared to a monodisperse system, and we show that polydispersity has very little effect on the intercalation behavior yet leads to capacity loss during the conversion reaction. We provide the code as an open-source Python Battery Mathematical Modelling (PyBaMM) model that can be used to identify performance limitations for other conversion cathode materials.

\end{abstract}


\end{frontmatter}

\section{Introduction}

\begin{figure*}
    \centering
    {
        \phantomsubcaption\label{fig:methods:OCV}
        \phantomsubcaption\label{fig:methods:diagram}
    }
    \includegraphics[width=\linewidth]{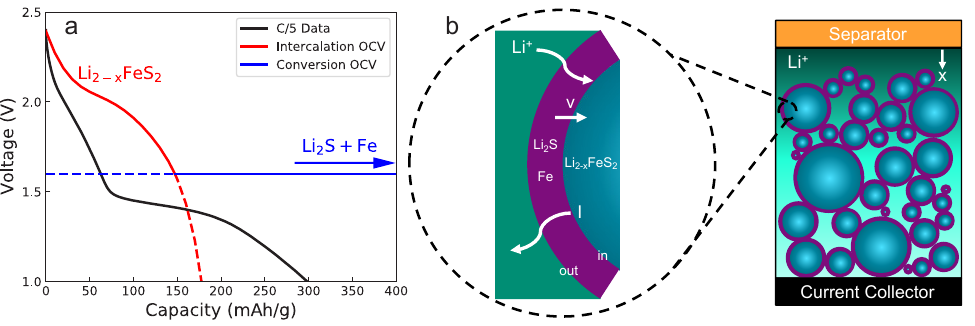}
    \caption{Illustration of the coupled intercalation-conversion reactions of \ce{FeS2}. (a) Characteristic C/5 discharge data and open circuit voltage (OCV) for the two-step intercalation-conversion reaction. (b) Diagram of half-cell P2D model showing (blue) the intercalated active material core and (purple) the converted shell of \ce{Li2S} and \ce{Fe}. The ionic lithium reacts at the reaction surface (interface between active material core and converted shell), and the current travels through the shell to the conductive binder (not shown) to eventually reach the current collector.}
    \label{fig:methods}
\end{figure*}

Due to the continually increasing demand for improved energy and power density in lithium-ion batteries, researchers have begun exploring ``next-generation'' cathode materials that offer theoretical capacities superior to existing materials \cite{Zhang2018,Wu2020}. While commercialized intercalation chemistries offer excellent cycle stability, the finite number of lithium storage sites in the host crystal structure results in limited capacity. In contrast, conversion-type electrodes undergo bond breakage and electrochemical transformation into new phases upon (de)lithiation yielding much higher capacities \cite{Kim2018,Yu2018,Wu2017}. One example of a conversion cathode material is \ce{FeS2} which undergoes a four-electron redox reaction and is of particular interest due to its abundance, low cost, and low toxicity \cite{Yersak2012,Rickard2007}. The reaction mechanism for \ce{FeS2} is currently debated, but the predominant presence of \ce{Li2S} at the end of discharge is universally agreed upon \cite{Zou2020,Zhang2015}. The most widely accepted mechanism for the rechargeable reaction pathway under a limited window (\SIrange{2.4}{1.0}{V}) is described as a two-step intercalation-conversion reaction:
\begin{align}
    \ce{
        Li_{2-n}FeS2 + n Li+ + n e- & <=> Li2FeS2, \label{eq:int_rxn}\\
        Li2FeS2 + 2Li+ + 2e- & <=> 2Li2S + Fe. \label{eq:conv_rxn}
    }
\end{align}
In this work, we consider the lower conversion and intercalation reactions of \ce{FeS2} by imposing a \SI{2.4}{V} window cutoff. Starting from \SI{2.4}{V}, lithium intercalates \ce{Li_{2-n}FeS2} to form \ce{Li2FeS2} which subsequently converts to form a mixture of \ce{Li2S} and \ce{Fe} \cite{Yersak2012,Fong1989}. This behavior manifests in the discharge curve in \cref{fig:methods:OCV} as two distinct regions, corresponding to the two distinct reactions.

Using the assumed two-step reaction pathway, several questions and areas for optimization still exist before \ce{FeS2} could have widespread use as a cathode material in secondary batteries. Particularly, the experimentally obtained capacities are far less than the theoretical capacity of \SI{894}{mAh/g} \cite{Yersak2012}, \ce{FeS2} shows poor cyclability when cycled over a wide voltage range \cite{Cabana2010}, and cells exhibit significant capacity loss when moving to faster rates \cite{Zou2020}. Several methods have been proposed to improve the overall performance of \ce{FeS2} including suppressing polysulfide dissolution by limiting the voltage window \cite{Schorr2021}, physically confining polysulfides to prevent shuttling \cite{Cui2022}, and limiting the debilitating effects of particle expansion by pressurizing the system \cite{Ashby2022} or designing nanonetworks to accommodate expansion \cite{Su2018}. Nevertheless, to unlock the full capacity of \ce{FeS2}, one must be able to isolate the exact reasons for the performance limitations, which can be difficult experimentally due to the complex reaction pathway and presence of multiple species during cycling.

Qualitatively, the two-step reaction mechanism can be envisioned as shown in \cref{fig:methods:diagram} for a half-cell. During discharge, a variety of transport mechanisms contribute to the overall cell performance. Particularly, lithium ions are transported through the electrolyte (green gradient), originating from the separator (yellow). The ionic lithium subsequently reacts at the surface of the particles to initially undergo an intercalation reaction where the intercalated lithium must then distribute throughout the active material particles (blue gradient). As the discharge proceeds to near completion, the overpotential will eventually reach a threshold where the conversion reaction will become favorable and initiate, and the intercalated \ce{Li2FeS2} begins to convert to a shell of \ce{Li2S} and \ce{Fe} (purple), rather than further intercalate into the particles. The shell region is assumed to allow lithium ions to diffuse through it to the reaction surface (blue-purple interface) as well as electrical current out of it to the conductive binder (not shown). The conversion reaction proceeds until either the intercalated core is completely consumed or the cell reaches a cutoff voltage. To balance the charge for the reaction, current is transported from the reaction surface to the current collector (black) through conductive binder between the particles (not shown).

The need to understand leading loss mechanisms for an electrochemical cell has warranted significant interest in electrochemical modeling. While mesoscale models are able to resolve complex behavior over various length scales and couple vast physical phenomena occurring within the cell, they typically require significant computing resources and are not efficient for testing wide sets of parameter ranges or experimental conditions \cite{Roberts2016}. In contrast, continuum models, such as the single particle model (SPM) or pseudo-two-dimensional (P2D) model, are widely used due to their robustness and ability to capture macroscopic effects accurately \cite{Zhang2000,Doyle1993,Santhanagopalan2006,Ramadesigan2012}. The P2D model, also referred to as the Doyle-Fuller-Newman (DFN) model, considers ionic and electrical transport throughout the electrode in combination with species transport throughout the active material particles to predict the overall behavior of the cell \cite{Doyle1993,Marquis2019}. Despite its widespread use, the model has primarily been applied to intercalation materials. Several authors have attempted to extend continuum models to account for phase change electrode materials \cite{Srinivasan2004,Subramanian2000,Jagannathan2009,Zhu2010,Prada2012}. However, these attempts typically require assumptions such as a monodisperse system, pseudo-steady-state for the solid-state diffusion, or neglect transport equations in the electrolyte. As such, there remains significant interest in development of an accurate, robust, and high-throughput model for conversion cathode materials, such as \ce{FeS2}, which can account for complexities such as polydispersity.

In this paper, we develop a P2D framework for polydisperse conversion cathode materials by using a core-shell model to represent the partially converted active material, with pseudo-steady-state solutions for ionic and electrical transport used throughout the converted shell material. Focusing on the lower conversion and intercalation reactions of \ce{FeS2} (\SIrange{2.4}{1.0}{V}), we fit our model to experimental lithiation data for \ce{FeS2} to identify the leading loss mechanisms in the cell. Our results indicate that for slow rates (C/20), lithium transport limitations in the active material are the leading loss mechanism in the intercalation regime, and poor electrical conductivity in the converted \ce{Li2S} and \ce{Fe} mixture is the leading loss mechanism in the conversion regime. At faster rates (C/5), we show that the intercalation and conversion reactions become mixed and additional transport limitations associated with electrode and ionic transport polarization become significant. We also assess the role of polydispersity by comparing our results to a monodisperse system and show that the intercalation behavior of the cell can be represented as monodisperse particles but that increasing polydispersity results in lower predicted maximum capacities in the conversion regime. Finally, we consider the role of ionic transport limitations in the converted shell and show that incorporating significant ionic transport limitations leads to a poor agreement with the experimental data over varying discharge rates. Although applied here to \ce{FeS2}, the conversion P2D model used throughout this work is robust and may be applied to other conversion cathode materials to better understand the performance limitations.

\section{Methods}
\subsection{Conversion Model Development}

\begin{table*}
    \centering
    \caption{Differential equations that comprise the model.}
    \label{tab:eqs}
    \small
    \renewcommand{\arraystretch}{2}
    \begin{tabular}{@{}lML@{}}
        \toprule\addlinespace[-1ex]
        Variable & \text{Equation}  \\[-1ex] 
        \midrule
        
        Solid potential ($V_s$) & 
        \frac{\partial}{\partial x}\left[\left(1-\epsilon\right)^b\kappa_\mathrm{eff}\frac{\partial V_s}{\partial x}\right]=aFj &
        eq:vs  \\
        Solid \ce{Li} concentration ($C_{\ce{Li}}$) & 
        \frac{\partial C_{\ce{Li}}}{\partial t}=-\frac{1}{r^2}\frac{\partial}{\partial r}\left(r^2\frac{D_{\ce{Li}}FC_{\ce{Li}}}{RT}\frac{\partial V_\mathrm{eq}}{\partial r}\right) & 
        eq:cs \\
        
        Electrolyte potential ($V_l$) & 
        \frac{\partial}{\partial x}\left[\epsilon^b\kappa_l\left(\frac{\partial V_l}{\partial x}-\left(1-t_+\right)\frac{2RT}{FC_{\ce{Li+}}}\frac{\partial C_{\ce{Li+}}}{\partial x}\right)\right]=-aFj & 
        eq:vl \\
        
        Electrolyte concentration ($C_{\ce{Li+}}$) & 
        \epsilon\frac{\partial C_{\ce{Li+}}}{\partial t}=\frac{\partial}{\partial x}\left(\epsilon^b D_{\ce{Li+}}\frac{\partial C_{\ce{Li+}}}{\partial x}\right)+a\left(1-t_+\right)j &
        eq:cl \\
        
        Shell thickness ($\delta$) & 
        \frac{\partial\delta}{\partial t}=-\frac{j_\mathrm{conv}M_\mathrm{FeS_2}}{\rho\left(4-2\frac{C_{\ce{Li}}}{C_\mathrm{max}}\right)} &
        eq:delta \\
        
        \bottomrule
    \end{tabular}
\end{table*}

\begin{table*}
    \centering
    \caption{Initial and boundary conditions applied to the governing equations shown in \cref{tab:eqs}.}
    \label{tab:bcs}
    \small
    \renewcommand{\arraystretch}{2}
    \begin{tabular}{@{}MMMM@{}}
        \toprule\addlinespace[-1ex]
        \text{Variable} & \text{Initial Condition} & \multicolumn{2}{c}{Boundary Conditions} \\[-1ex]
        \cmidrule{3-4} \addlinespace[-1ex]
        & & x=0 & x=L  \\[-1ex]
        \midrule 
        
        V_s & 
        V_s=\SI{2.4}{V}^* & 
        \frac{\partial V_s}{\partial x}=0 &
        \frac{\partial V_s}{\partial x}=-\frac{i_\mathrm{cell}}{\left(1-\epsilon\right)^b\kappa_\mathrm{eff}} \\
        
        C_{\ce{Li}} &
        C_{\ce{Li}}=0.6 C_\mathrm{max} & 
        \frac{\partial C_{\ce{Li}}}{\partial r}^\dagger=0 &
        \frac{D_{\ce{Li}}FC_{\ce{Li}}}{RT}\frac{\partial V_\mathrm{eq}}{\partial r}^\ddagger=j_\mathrm{int} \\
        
        V_l & 
        V_l=0^* & 
        V_l=0 &
        \frac{\partial V_l}{\partial x}=0 \\
        
        C_{\ce{Li+}} &
        C_{\ce{Li+}}=C_0 &
        
        C_{\ce{Li+}}=C_0 & 
        \frac{\partial C_{\ce{Li+}}}{\partial x}=0 \\
        
        \delta &
        \delta=0 & 
        \text{---} &
        \text{---} \\
        
        \bottomrule
        \multicolumn{4}{l}{$^*$ Initial guess} \\[-1ex]
        \multicolumn{4}{l}{$^\dagger$ Evaluated at $r=0$ instead of $x=0$} \\[-1ex]
        \multicolumn{4}{l}{$^\ddagger$ Evaluated at $r=R_p-\delta$ instead of $x=L$} \\
    \end{tabular} 
\end{table*}

The P2D model used throughout this work is based off the half-cell DFN model, modified with an additional ordinary differential equation added to represent the shell thickness evolution during the conversion reaction and mechanisms to transport through that shell (see \cref{fig:methods:diagram}). This model includes domains in which transport is considered: throughout the electrode and throughout the active particle core. The governing equations for our model are shown in \cref{tab:eqs}, and the boundary and initial conditions are provided in \cref{tab:bcs} \cite{Doyle1993,Marquis2019}. A list of all parameters and variables with their respective definitions may be found in the nomenclature tables \cref{tab:symbol,tab:subscript} at the end of the text. 

Five differential equations are solved simultaneously: electrical current transport through the solid phase network (active material particles and conductive binder domain, \cref{eq:vs}), intercalated lithium transport radially within the active material particle core (\cref{eq:cs}), current transport in the electrolyte (\cref{eq:vl}), ionic transport in the electrolyte (\cref{eq:cl}), and the active material particle shell thickness (\cref{eq:delta}). With the exception of \cref{eq:cs}, which is solved in the radial particle domain, all equations are solved throughout the linear electrode domain (labeled $x$ in \cref{fig:methods:diagram}). Note that in contrast to the traditional DFN model, diffusion throughout the intercalated active material is assumed to be non-ideal (i.e. follow concentrated solution theory) where the lithium flux is driven by gradients in the equilibrium potential $V_\mathrm{eq}$ rather than gradients in the lithium concentration $C_{\ce{Li}}$.

The total ionic lithium flux due to consumption through electrochemical reactions $j$ is the summation of two independent fluxes from the intercalation reaction $j_\mathrm{int}$ and the conversion reaction $j_\mathrm{conv}$. In the electrode transport equations, this surface flux is multiplied by the specific active surface area $a$ to obtain the volumetric flux. The conversion reaction and intercalation reaction each obey separate Butler-Volmer equations:
\begin{equation}
    j_i=\frac{i_{0,i}}{F}\left[\exp{\left(\frac{\alpha z F\eta_i}{RT}\right)}-\exp{\left(-\frac{\alpha z F\eta_i}{RT}\right)}\right],
    \label{eq:BV}
\end{equation}
where $i_{0}$ is the exchange current density, $\alpha$ is the charge transfer coefficient, $z$ is the number of electrons in the reaction, and the subscript $i$ corresponds to either the intercalation \textit{int} or conversion reaction \textit{conv}. Note that both reactions considered throughout this work are analyzed on a per-electron basis, so $z$ is fixed at 1 for our analysis. The overpotential $\eta_i$ is defined as:
\begin{equation}
    \eta_i=V_{s,\mathrm{in}}-V_{l,\mathrm{in}}-V_{\mathrm{eq},i}\label{eq:overpotential}
\end{equation}
where $V_{s,\mathrm{in}}$, $V_{l,\mathrm{in}}$, and $V_{\mathrm{eq},i}$ are the solid, liquid, and equilibrium potentials, respectively, at the reaction surface. The intercalating lithium flux into the particle at the reaction surface (the second boundary condition for $C_{\ce{Li}}$ in \cref{tab:bcs}) is only governed by the intercalation reaction \cref{eq:int_rxn}, where the exchange current density depends on both the intercalated lithium concentration $C_{\ce{Li}}$ and the ionic lithium concentration at the reaction surface $C_{\ce{Li+},\mathrm{in}}$:
\begin{equation}
    i_{0,\mathrm{int}}=k_\mathrm{int}F\left(C_{\ce{Li}}\right)^\alpha \left(C_{\ce{Li+},\mathrm{in}}\right)^\alpha\left(C_\mathrm{max}-C_{\ce{Li}}\right)^\alpha
\end{equation}
where $k_\mathrm{int}$ is a rate constant for the intercalation reaction and $C_\mathrm{max}$ represents the maximum lithium concentration in the active particles. The intercalation equilibrium potential is an empirical function determined from experimental Galvanostatic intermittent titration technique (GITT) measurements and depends on the lithium concentration, as discussed in the Supplemental Information \cite{Horner2021}.

The conversion reaction \cref{eq:conv_rxn} also contributes a source for the current density through a separate Butler-Volmer expression. However, unlike the intercalation reaction, the conversion reaction does not provide a lithium source into the active material. Instead, the reaction rate is linked to the shell radius evolution \cref{eq:delta} \cite{Vijayasekaran2006,Srinivasan2004}. As the conversion reaction proceeds to completion, the intercalated active material core will be converted to a mixture of \ce{Li2S} and \ce{Fe} (\cref{eq:conv_rxn}). This consumption drives the reaction surface (the interface between the intercalated core and converted shell material) inward, thus resulting in a shrinking core mechanism. The specific form used for \cref{eq:delta} ensures conservation of species and converts the molar reaction rate to a volumetric flux.

For the conversion reaction, we use a second order kinetic rate expression that depends only on the ionic lithium concentration at the reaction surface:
\begin{equation}
    i_{0,\mathrm{conv}}=k_\mathrm{conv}F\left(C_{\ce{Li+},\mathrm{in}}\right)^2 .
\end{equation}
The second order reaction with respect to the ionic lithium concentration was found to represent the experimental data over different discharge rates marginally better than a value of 0.5 for the charge transfer coefficient (Supplemental Information). We use a constant equilibrium potential for the conversion reaction, $V_\mathrm{eq,conv}=\SI{1.6}{V}$. Note that throughout this work, we only consider the discharge behavior of \ce{FeS2} to mitigate capacity fade mechanisms across cycles and identify the intra-cycle polarization mechanisms. As such, a conditional is placed on the voltage to prevent the conversion reaction from proceeding in the reverse direction above the OCV. If one were interested in simulating the charge behavior, this conditional would need to be replaced with a conditional on the shell thickness as it approaches zero. The intercalation portion of our model is already equipped for reversibility and does not need modification for charge simulations. 

The presence of the converted shell on the outside of the active material causes additional transport losses with respect to the ionic lithium concentration and the current in the electrode and electrolyte. For simplicity, we assume that the transport through the shell is in pseudo-steady-state, meaning that the steady-state solution can be used to explicitly calculate the electrode voltage $V_{s,\mathrm{in}}$, electrolyte voltage $V_{l,\mathrm{in}}$, and ionic lithium concentration $C_{\ce{Li+},\mathrm{in}}$ at the reaction surface (i.e. the surface between the active material core and the converted shell):
\begin{align}
    V_{s,\mathrm{in}}&=V_{s,\mathrm{out}}-\frac{jFR_c^2}{\kappa_\mathrm{shell}}\left(\frac{1}{R_c}-\frac{1}{R_p}\right) \label{eq:solidloss} \\
    V_{l,\mathrm{in}}&=V_{l,\mathrm{out}}+\ln\left(\frac{C_\mathrm{\ce{Li+},in}}{C_\mathrm{\ce{Li+},out}}\right)\frac{RT}{F} \label{eq:elecloss} \\
    C_{\ce{Li+},\mathrm{in}}&=C_\mathrm{\ce{Li+},out}+\frac{jR_c^2}{2D_\mathrm{shell}}\left(\frac{1}{R_c}-\frac{1}{R_p}\right) \label{eq:concloss}
\end{align}
where $j$ is the summation of the ionic flux from the intercalation and conversion reaction, $\kappa_\mathrm{shell}$ is the solid-phase electrical conductivity in the shell, $D_\mathrm{shell}$ is the ionic diffusivity through the shell, $R_p$ is the total radius of the particle (i.e. including the active material core and the converted shell), and $R_c=R_p-\delta$ is the radius of the active material core. The in and out subscripts indicate whether the quantity is at the reaction surface (blue-purple interface in \cref{fig:methods:diagram}) or at the outer surface of the converted shell (green-purple interface in \cref{fig:methods:diagram}). The outer surface quantities are used in the electrode transport equations and the reaction surface quantities are used in the kinetic expressions. We provide the derivations for these equations in the Supplemental Information.

Finally, the growth of the shell layer affects the effective solid phase electrical conductivity $\kappa_\mathrm{eff}$ through a Bruggeman effective medium approximation:
\begin{equation}
    0=\sum_i{\phi_i\frac{\kappa_i-\kappa_\mathrm{eff}}{\kappa_i+2\kappa_\mathrm{eff}}}
    \label{eq:cond}
\end{equation}
where $\phi_i$ is the volume fraction of the particular phase and the index $i$ corresponds to either the active material, converted shell, or carbon-binder domain (CBD) phase \cite{Bruggeman1935}. Note that $\kappa_\mathrm{eff}$ represents the effective conductivity of the solid material, and the effect of the electrolyte is incorporated through the porosity dependent prefactor in \cref{eq:vs}.

Several assumptions underlie our model that should be noted. Particularly, as the model is an extension of the DFN model, the assumptions from the DFN model also apply here. Namely, the electrode particles are assumed to be spherical, the electrolyte is treated as a single phase, and approximations are used to describe the effective properties across the electrode and electrolyte. Additionally, as mentioned previously, we assume the transport through the converted shell layer to be in pseudo-steady-state. We are also using only a half-cell model, so effects associated with the anode and separator are neglected, and we assume that an infinite lithium-ion source is present at the separator-cathode interface.

\subsection{Incorporating Polydispersity}\label{methods-polydispersity}

Our previous work on analyzing GITT measurements for \ce{FeS2} \cite{Horner2021} suggested that for intercalation materials, a polydisperse system of particles could be represented as a system of monodisperse particles with an effective radius equal to three times the ratio of the total volume to total active surface area. However, for conversion materials this approach is insufficient as the core radius is varying with extent of reaction. We anticipate transport through the shell will be the primary polarization loss mechanism and that the surface reaction flux is approximately the same for all particle sizes, and thus the shell thickness is approximately equal for all particle sizes at a particular distance across the electrode. This assumption is supported by our previous analysis of the intercalation regime of \ce{FeS2} \cite{Horner2021} and will be explored in \cref{sec:polar}. The increasing shell thickness leads to the complete consumption of smaller particles prior to that of larger particles and therefore a change in the effective particle radius and available specific surface area of the active materials with the extent of the conversion reaction.

\begin{figure*}
    \centering
    {
        \phantomsubcaption\label{fig:polydispersity-methods:PSD}
        \phantomsubcaption\label{fig:polydispersity-methods:radius}
        \phantomsubcaption\label{fig:polydispersity-methods:area}
    }
    \includegraphics[width=\linewidth]{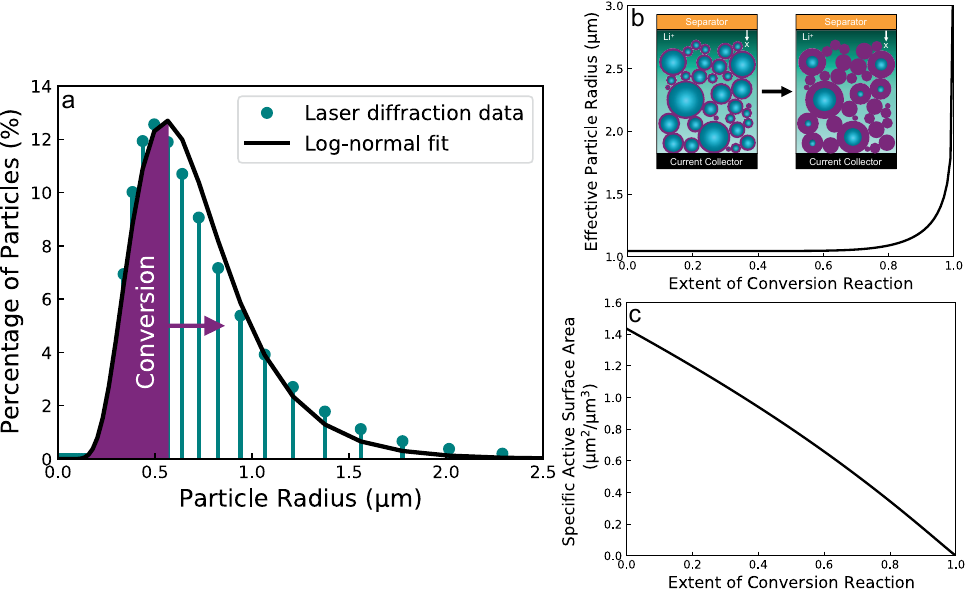}
    \caption{Particle size polydispersity evolution. (a) Particle size distribution (PSD) obtained from laser diffraction measurements with log-normal fit \cite{Horner2021}. Purple shaded region represents the conversion reaction progressing which consumes the smaller particles first. (b) Effective particle radius as a function of extent of conversion reaction. Inset shows the conversion reaction progressing to leave only the larger particles active. (c) Specific active surface area $a$ as a function of extent of conversion reaction.}
    \label{fig:polydispersity-methods}
\end{figure*}

To address this, we implement a variable particle radius approach, in which we assume the shell thickness for all particles at a specific distance across the electrode to be equal. As the conversion reaction progresses, the smaller particles will convert entirely leaving only the larger particles to continue participating in reactions. \Cref{fig:polydispersity-methods:PSD} shows the measured particle size distribution of our active material. At an extent of reaction of 0.84, all particles with radius \SI{<0.565}{\um} have been fully consumed, as represented by the purple shaded region. Therefore, only the particles with radius \SI{>0.565}{\um} participate in further chemical conversion. As a result, the effective particle radius $R_{p,\mathrm{eff}}$ will increase, and the specific active surface area $a$, which converts the surface lithium flux into a volumetric flux, will decrease with increasing shell thickness. These results are shown as a function of the extent of conversion reaction in \cref{fig:polydispersity-methods:radius,fig:polydispersity-methods:area}. We provide the analytical expressions for these quantities in the Supplemental Information.

Note that the shrinking core mechanism is still captured through the shell thickness evolution equation (\cref{eq:delta}). However, the outer particle radius in \cref{eq:solidloss,eq:elecloss,eq:concloss} is now variable and increasing with the extent of conversion reaction. To ensure we are conserving mass in the system, we monitored the total moles of lithium during discharge. The mass imbalance for our study was ~0.3\% which is likely due to a combination of numerical issues and our assumption that the intraparticle lithium profile is not affected by the shrinking core. Nevertheless, this change in mass was small, and we do not anticipate that it significantly affects the results.

\subsection{Model Implementation}

The model is implemented using the open-source Python Battery Mathematical Modelling (PyBaMM) software \cite{Andersson2018,Harris2020,Sulzer2020}. The model contains two spatial dimensions: the distance across the electrode from separator to collector $x$ (for \cref{eq:vs,eq:vl,eq:cl}) and the radial distance from the particle center $r$ (for \cref{eq:cs}). Even though \cref{eq:cs,eq:delta} do not have derivatives with respect to $x$, $C_\mathrm{Li}$ and $\delta$ do vary with $x$. All differential equations listed in \cref{tab:eqs} are solved simultaneously, and additional algebraic equations corresponding to the current density for both reactions (\cref{eq:BV}) and the effective conductivity (\cref{eq:cond}) are also solved numerically. Note that the current density equations must be solved numerically as the solid voltage at the reaction surface depends on the ionic lithium flux (\cref{eq:solidloss}).

\begin{table*}
    \centering
    \caption{\ce{FeS2} parameters used in the model.}
    \small
    \begin{tabular}{@{}llSS[table-text-alignment=left]l@{}}
        \toprule
        Parameter & Symbol & {Value} & {Unit} & Source \\ \midrule
        Bruggeman coefficient & $b$ & 1.5 & & --\\
        Initial electrolyte concentration & $C_0$ & 1e3 & \si{mol/m^3} & --\\
        Electrode solid-state diffusivity & $D_{\ce{Li}}$ & 8.74e-19 & \si{m^2/s} & \cite{Horner2021} \\
        Electrolyte ionic diffusivity  & $D_{\ce{Li+}}$ & 6e-12 & \si{m^2/s} & \cite{Castiglione2011}\\
        Shell ionic diffusivity & $D_\mathrm{shell}$ & 6e-13 & \si{m^2/s} & Estimated*\\
        Faraday constant & $F$ & 96485 & \si{s.A/mol} & --\\
        Conversion reaction rate constant & $k_\mathrm{conv}$ & 2e-14 & \si{m^4/mol.s} & GITT\\
        Intercalation reaction rate constant & $k_\mathrm{int}$ & 5.73e-12 & \si{m^{2.5}/mol^{0.5}.s} & \cite{Horner2021} \\
        Electrode thickness & $L$ & 30 & \si{\micro m} & Estimated\\
        \ce{FeS2} molar mass & $M_\mathrm{FeS_2}$ & 0.120 & \si{kg/mol} & \cite{Pyrite} \\
        Ideal gas constant & $R$ & 8.314 & \si{J/mol.K} & --\\
        Temperature & $T$ & 298 & \si{K} & --\\
        Transference number & $t_+$ & 0.16  & & \cite{Yan2021}\\
        Charge transfer coefficient & $\alpha$ & 0.5 & & Assumed \\
        Number of electrons & $z$ & 1 & & -- \\
        Electrode porosity & $\epsilon$ & 0.5 & & Estimated\\
        Binder conductivity & $\kappa_\mathrm{CBD}$ & 1 & \si{S/m} & Estimated*\\
        Electrolyte conductivity & $\kappa_l$ & 0.15 & \si{S/m} & \cite{Nadherna2011}\\
        Electrode conductivity & $\kappa_s$ & 0.001 & \si{S/m} & Estimated*\\
        Shell conductivity & $\kappa_\mathrm{shell}$ & 1e-7 & \si{S/m} & Fit\\
        \ce{FeS2} density & $\rho$ & 5010 & \si{kg/m^3} & \cite{Pyrite} \\
        PSD (\si{\um}) $\ln()$ mean & $\mu$ & -0.411 & & \cite{Horner2021} \\
        PSD (\si{\um}) $\ln()$ standard deviation & $\sigma$ & 0.427 & & \cite{Horner2021} \\
        \bottomrule
        \multicolumn{5}{l}{{\scriptsize * Indicates the estimated value had no significant impact on the results within several orders of magnitude.}}
    \end{tabular}
    \label{tab:params}
\end{table*}

Relevant parameters for the model are provided in \cref{tab:params}. When available, parameter values from literature were used. Unfortunately, we were unable to find values for the intercalated active material conductivity, the shell conductivity, and the shell ionic diffusivity. The intercalated active material conductivity was found to have minimal effect on the results due to the presence of the CBD. Here, we estimate the primary losses through the shell to be due to poor electrical conductivity with negligible ionic diffusive losses. This leaves the shell conductivity as a single fit parameter, which was fit using the SciPy curve fit function to the C/5 experimental data. The validity of this assumption and incorporation of non-negligible ionic diffusion losses in the shell will be discussed in \cref{ionic-limitations}.

The model is initialized with the shell thickness and electrolyte voltage set to zero. The ionic lithium concentration starts at \SI{1}{M}, and the solid voltage starts at \SI{2.4}{V} -- both values that are dictated by the experimental setup. The solid lithium concentration starts at a value corresponding to \ce{Li_{1.2}FeS2} (\SI{50100}{mol/m^3}), which is the delithiated state of the intercalation reaction\cite{Fong1989,Son2013}. Simulations are run until the voltage at the collector reaches a cutoff voltage of \SI{1}{V}. 

\subsection{Experimental Methods}
\subsubsection{Materials Preparation}

\ce{FeS2} powder was purchased from Sigma and subsequently ball milled for a continuous 6 hours at 1000 rpm in a Fritsch Pulverisette 7 Premium Line Planetary Micro Mill using a \SI{20}{mL} stainless steel Fritsch grinding bowl and \SI{3}{mm} stainless steel Fritsch media (equal mass to the sample being milled). \ce{FeS2} slurry electrodes were made by mixing the ball milled \ce{FeS2} powder, Super P (Alfa Aesar), and poly(vinylidene difluoride)(Kynar Flex 2801) dissolved in N-methyl-2-pyrrolidone (Sigma-Aldrich) in a 80:10:10 weight ratio, respectively. Contents were mixed using a mortar and pestle and subsequently doctor bladed onto a carbon-coated aluminum foil current collector (MTI Corp.). Slurries were dried in a vacuum oven at \SI{120}{\celsius} for at least 12 hours and then directly transferred into an argon glovebox (\SI{<1}{ppm} \ce{O2},  \ce{H2O}) for coin cell assembly. \ce{FeS2} electrodes used in this study had a diameter of \SI{0.375}{in}, and the mass loading was \SIrange{1}{1.5}{mg/cm^2}.

\subsubsection{Materials Characterization}

To determine the particle size distribution of \ce{FeS2} particles, laser diffraction measurements were carried out using a Malvern Mastersizer 3000 instrument with a dispersion unit accessory. The dispersing solvent used was Vertrel XF, the refractive index was set at 3.08, and the absorption index at 0.01 before running the measurements. The solvent was first added to the dispersion unit accessory, after which 1-5 mg of sample was added in dry form. Between sample measurements, fresh solvent was flushed through the system to prevent cross-contamination and faulty measurements.

\subsubsection{Electrochemical Characterization}

Two electrode coin cells were assembled using a Celgard H1409 trilayer separator with \SI{80}{\micro L} of \SI{1}{M} \ce{LiFSI} \ce{PYR14TFSI} ionic liquid electrolyte and a \SI{0.5}{in} diameter lithium metal counter electrode. The same electrode was used as the reference electrode, and we assumed the plating/stripping overpotential for the lithium metal to be negligible. An upper cutoff voltage of \SI{2.4}{V} was applied to avoid the upper conversion reaction in \ce{FeS2} and mitigate polysulfide shuttling and dissolution \cite{Schorr2021}. Therefore, the \SIrange{2.4}{1.0}{V} window used in this study captures the lower conversion reaction and intercalation region. The discharge rates used in this study (C/20 -- \SI{0.34}{mA/cm^2}, C/10 -- \SI{0.67}{mA/cm^2}, and C/5 -- \SI{1.34}{mA/cm^2}) were based off the full four-electron process capacity of \ce{FeS2} (\SI{894}{mAh/g}) GITT measurements used for OCV extrapolation were obtained using a 20 minute pulse at C/20 followed by a 4 hour rest step from \SIrange{2.4}{1.0}{V}.

\section{Results and Discussion}
\subsection{Nominal Behavior at C/5}

\begin{figure*}
    \centering
    {
        \phantomsubcaption\label{fig:C5:C5}
        \phantomsubcaption\label{fig:C5:contour}
        \phantomsubcaption\label{fig:C5:conversion}
        \phantomsubcaption\label{fig:C5:shell}
        \phantomsubcaption\label{fig:C5:lithium}
        \phantomsubcaption\label{fig:C5:solid}
        \phantomsubcaption\label{fig:C5:lithium-ion}
        \phantomsubcaption\label{fig:C5:electrolyte}
    }
    \includegraphics[width=\linewidth]{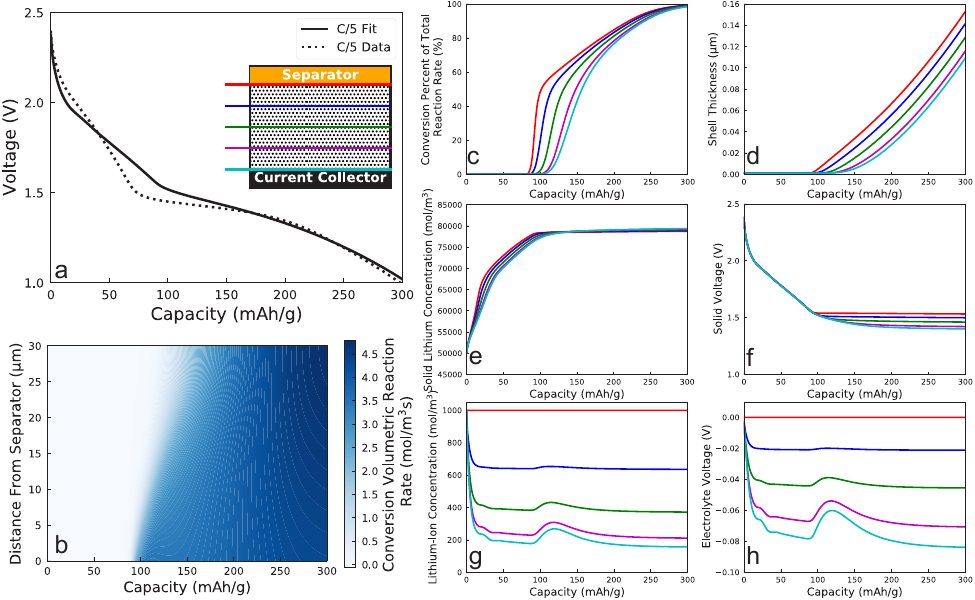}
    \caption{Model predictions and comparison to experimental data for a C/5 discharge rate. (a) C/5 experimental discharge data and model prediction. Inset shows a schematic of the cell with colored lines used in (c)-(h) marking their different locations across the cell. The texture within the cell reflects the porous electrode/electrolyte system with a thickness of \SI{30}{\micro m}. (b) Contour map showing the volumetric reaction rate for the conversion reaction as a function of the specific capacity and distance from the separator. (c) Percent of total reaction rate that corresponds to the conversion reaction, (d) shell thickness, (e) solid lithium concentration, (f) solid voltage, (g) lithium-ion concentration in the electrolyte phase, and (h) electrolyte voltage are all shown as a function of capacity and evaluated at the reaction surface.}
    \label{fig:C5}
\end{figure*}

We begin our results by exploring the model predictions for the highest experimental discharge data available, C/5, in \cref{fig:C5}. The model results are compared to experimental discharge data are shown in \cref{fig:C5:C5}. Additionally, we provide a contour map showing the transient progression of the conversion reaction throughout the electrode and several variables of interest at different distances from the separator (indicated as the line color as defined in the inset schematics of \cref{fig:C5:C5}) as a function of capacity. All depicted parameters are at the reaction surface (subscript \textit{in} in the model formulation). Additional results for C/10 and C/20 are provided in the Supplemental Information. 

The model prediction matches the qualitative behavior of the experimental discharge profile in \cref{fig:C5:C5} and the maximum capacity well. However, some deviation between the model prediction and experimental results can be seen primarily in the last half of the intercalation region. The model over-predicts the capacity for the intercalation region indicating that some loss mechanisms in this region are not fully captured. Due to the fact that we were able to match the intercalation region well in our previous work when the voltage was restricted between \SIrange{1.6}{2.4}{V} \cite{Horner2021}, the poorer agreement with this region shown in \cref{fig:C5:C5} may indicate that additional irreversible side reactions are occurring at lower voltages that lead to capacity loss, which is supported by literature \cite{Park2013,Evans2014}.

Model quantities within the solid phase of the electrode show comparatively little variation across the cell thickness. The reactions occur more quickly near the separator interface, as shown in the contour map for the conversion reaction (\cref{fig:C5:contour}). For the intercalation reaction, this leads to a slightly higher solid lithium concentration during the intercalation phase(\cref{fig:C5:lithium}). The lithium concentration then approaches an identical value across the cell which is still below the maximum concentration of \SI{83500}{mol/m^3}. The solid voltage (\cref{fig:C5:solid}) exhibits the opposite trend, where little variation is seen for the intercalation reaction at early capacities followed by variation across the cell during the conversion reaction. This is due to the high conductivity of the binder phase which helps to mitigate solid voltage polarization across the electrode. Mathematically, this can be shown through \cref{eq:vs} where $\kappa_\mathrm{eff}$ will be large, requiring the spatial derivatives to be small to maintain the equality. However, once the conversion reaction begins, the solid voltage is more dependent on the shell thickness, as electrons must cross the shell before reaching the conductive binder, and the shell thickness will vary across the cell as shown in \cref{fig:C5:shell}. \cref{fig:C5:solid}. This identifies the solid electrical transport through the shell as a leading loss mechanism in the conversion regime.

The electrolyte properties show much more significant variation across the cell as seen in \cref{fig:C5:lithium-ion,fig:C5:electrolyte}, which is due to ionic transport limitations that become especially pronounced at faster discharge rates. Upon initiation of the conversion reaction (around \SI{100}{mAh/g} for the C/5 results, \cref{fig:C5:conversion}), the model predicts a short transient increase in the lithium-ion concentration and the electrolyte voltage, most notably closer to the current collector, which arises due to the initiation of the conversion reaction once the overpotential becomes negative. The conversion reaction first occurs closest to the separator as can be seen in \cref{fig:C5:contour,fig:C5:conversion}, and the additional source of current requires less load from particles closer to the current collector. The decreased reaction rate at particles closer to the current collector leads to a build-up of lithium ions here which eventually relaxes as the conversion reaction propagates toward the current collector. The transient increase in the electrolyte voltage is due to the same effect, as the electrolyte voltage and lithium-ion concentration are inherently connected.

\subsection{Polarization Loss Mechanisms at Varying Discharge Rates}\label{sec:polar}

\begin{figure*}
    \centering
    {
        \phantomsubcaption\label{fig:crates:crates}
        \phantomsubcaption\label{fig:crates:C5losses}
        \phantomsubcaption\label{fig:crates:C10losses}
        \phantomsubcaption\label{fig:crates:C20losses}
        \phantomsubcaption\label{fig:crates:C5extents}
        \phantomsubcaption\label{fig:crates:C10extents}
        \phantomsubcaption\label{fig:crates:C20extents}
    }
    \includegraphics[width=\linewidth]{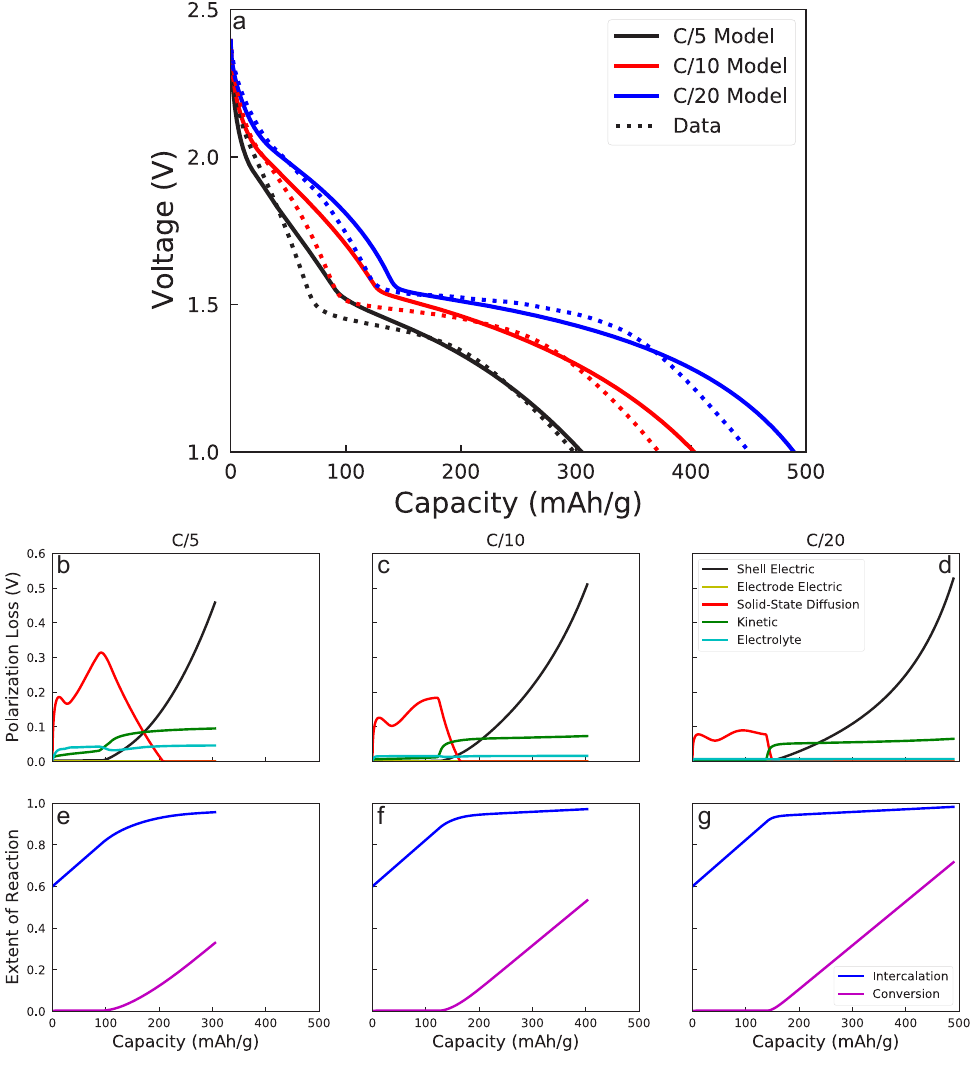}
    \caption{(a) Experimental discharge data and model predictions at C/5, C/10, and C/20. (b)-(d) Various polarization losses as a function of capacity for different discharge rates. (e)-(g) Extents of intercalation and conversion reactions as a function of capacity for different discharge rates.}
    \label{fig:crates}
\end{figure*}

In \cref{fig:crates:crates}, we demonstrate that our model performs well at multiple discharge rates of C/5, C/10, and C/20. Upon increasing the discharge rate, the data exhibit capacity loss both in the intercalation region and the conversion region. The model matches the capacity loss that arises when moving to faster discharge rates and several qualitative features of the discharge curve. However, the model fails to capture the exact quantitative behavior of the cell. The discrepancies are similar in magnitude to those seen for the C/5 results and are likely a result of model limitations discussed in the previous section.

To better understand the limitations of the cell when moving to faster discharge rates as well as at slow rates, we consider several loss mechanisms, depicted in \cref{fig:crates:C5losses,fig:crates:C10losses,fig:crates:C20losses}. Each of these losses represents a voltage polarization mechanism that causes the resulting cell voltage to be lower than the reaction's equilibrium potential. The exact definitions for the various losses shown in \cref{fig:crates:C5losses,fig:crates:C10losses,fig:crates:C20losses} are listed in \cref{tab:losses}.

\begin{table*}
    \centering
    \caption{Definitions of polarization losses depicted in \cref{fig:crates:C5losses,fig:crates:C10losses,fig:crates:C20losses}.}
    \label{tab:losses}
    \small
    \begin{tabular}{@{}p{0.15\linewidth}p{0.43\linewidth}M@{}}
        \toprule
        
        Loss \mbox{Mechanism} & 
        Description & 
        \text{Formula} \\ \midrule
        
        Shell electric & 
        Solid-phase electrical conductivity loss through the converted shell material & 
        \int_{0}^{L} \frac{ajF}{i_\mathrm{cell}}\left(V_{s,\mathrm{out}}-V_{s,\mathrm{in}}\right) \,dx \\
        
        Electrode electric & 
        Solid-phase electrical conductivity loss through the electrode & 
        \int_{0}^{L} \frac{ajF}{i_\mathrm{cell}}\left(V_\mathrm{cell}-V_{s,\mathrm{out}}\right) \,dx \\
        
        Solid-state diffusion & 
        Difference between reaction surface (in) OCV and OCV based on average lithium concentration within the active material. & 
        \int_{0}^{L} \frac{ajF}{i_\mathrm{cell}}\left[V_\mathrm{eq,in}-V_\mathrm{eq}\left(\overline{C_{\ce{Li}}}\right)\right] \,dx \\
        
        Kinetic & 
        Overpotential required to drive the reaction & 
        \int_{0}^{L} \frac{ajF}{i_\mathrm{cell}}\left(V_{s,\mathrm{in}}-V_{l,\mathrm{in}}-V_\mathrm{eq,in}\right) \,dx \\
        
        Electrolyte & 
        Change in electrolyte voltage across the electrode & 
        \int_{0}^{L} \frac{ajF}{i_\mathrm{cell}}V_{l,\mathrm{out}} \,dx \\
        
        \bottomrule
        \multicolumn{3}{l}{{\scriptsize The overbar represents the spatial mean across a particle.}}
    \end{tabular}
    
\end{table*}

At slow discharge rates (\cref{fig:crates:C20losses}), the intercalation regime occurs nearly exclusively until \SI{150}{mAh/g} (\cref{fig:crates:C20extents}) of capacity is used, and solid-state diffusion of lithium within the particles is the only polarization mechanism that is significant. However, when the conversion reaction begins, two loss mechanisms become important. Initially, the reaction is kinetically limited, with a polarization loss of \SI{\sim 0.6}{V} that remains relatively constant throughout the conversion regime. Very quickly after the beginning of the conversion regime, however, the particle shells, which have a very low electrical conductivity, quickly grow, leading to electrical transport losses from the inner reaction surface to the outer shell surface. This loss mechanism is unique to conversion cathodes (compared to pure intercalation materials) and grows exponentially as the conversion reaction proceeds, accounting for nearly all of the overall voltage drop late in the discharge and overall loss of usable capacity within the specified voltage window. Note that the solid-phase electrical losses through the electrode are insignificant compared to the other polarization mechanisms. This domination of shell electrical conductivity  losses validates our assumption of this phenomena discussed earlier (\cref{methods-polydispersity}).

At faster discharge rates (\cref{fig:crates:C10losses,fig:crates:C5losses}, the leading loss mechanisms are the same, but additional, non-negligible losses associated with ionic transport polarization arise. The additional losses explain some of the observed capacity loss at faster discharge rates, namely that the reaction occurs in a diffuse front that begins near the separator and slowly spreads and moves towards the collector which can also be seen in \cref{fig:C5:contour}. The significance of the shell electric polarization loss for all sampled C-rates also helps to exemplify the need for the adapted P2D model which accounts for an additional transport resistance across the shell. Had the standard P2D model been used for this material, the shell electric loss would not be accounted for, and one would obtain significantly worse predictions for a constant OCV in the conversion regime. The standard P2D model would predict much higher capacities as it would not account for the main loss mechanism in the conversion regime.

In addition to other loss mechanisms becoming relevant at faster discharge rates, the intercalation and conversion reaction also become more significantly mixed at faster discharge rates. Although the mixed reaction regime can be seen in the loss mechanisms, it may be seen more clearly in the extents of reactions shown in \cref{fig:crates:C5extents,fig:crates:C10extents,fig:crates:C20extents} as a function of capacity. At slow discharge rates, the reactions are distinct with the intercalation reaction going to almost full completion before the conversion reaction begins. However, at faster discharge rates, the reactions are happening simultaneously which results in neither going to full completion.

\subsection{Polydispersity}

\begin{figure*}
    {
        \phantomsubcaption\label{fig:polydispersity:results}
        \phantomsubcaption\label{fig:polydispersity:losses}
    }
    \includegraphics[width=\linewidth]{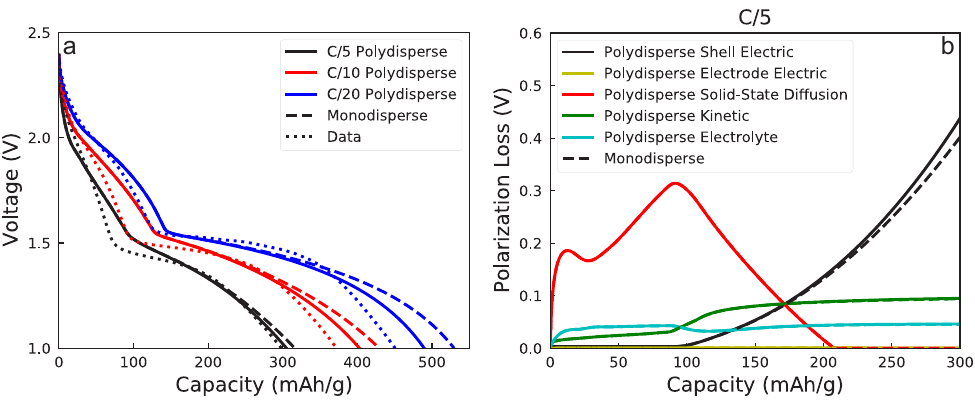}
    \caption{(a) Comparison of (solid lines) polydisperse and (dashed lines) monodisperse simulations at different discharge rates. Experimental data are shown as dotted lines. (b) Leading loss mechanisms for (solid lines) polydisperse and (dashed lines) monodisperse simulations at C/5.}
    \label{fig:polydispersity}
\end{figure*}

All results presented thus far have been for a polydisperse system of particles using the methodology for incorporating polydispersity described in \cref{methods-polydispersity}. To assess the impact that polydispersity has on the results, we compare the polydisperse simulation results to a monodisperse system with radius equal to three times the initial total volume to total surface area ratio of the active material ($R_p=\SI{1.045}{\micro m}$). The comparison is shown in \cref{fig:polydispersity:results}.

Similar to our previous results on GITT measurements \cite{Horner2021}, the polydisperse and monodisperse results align in the intercalation regime, as the effective particle radius equal to three times the total volume to total surface area ratio maintains the surface flux quantities for the electrode. However, in the conversion regime, incorporating polydispersity leads to a lower calculated capacity. The capacity loss is due to the increasing effective particle radius and the faster decrease in specific active surface area for the polydisperse system (\cref{fig:polydispersity-methods:radius,fig:polydispersity-methods:area}) when compared to the monodisperse results. As shown through the leading loss mechanisms presented in \cref{fig:polydispersity:losses}, the primary difference arises through the shell electric polarization. For the polydisperse case, the small particles get entirely consumed early on leaving only the larger particles, and thus increasing the shell thickness and transport losses through the shell as the reaction proceeds to completion.

\subsection{Shell Ionic Transport Limitations}\label{ionic-limitations}

\begin{figure}
    \centering
    \includegraphics[width=3in]{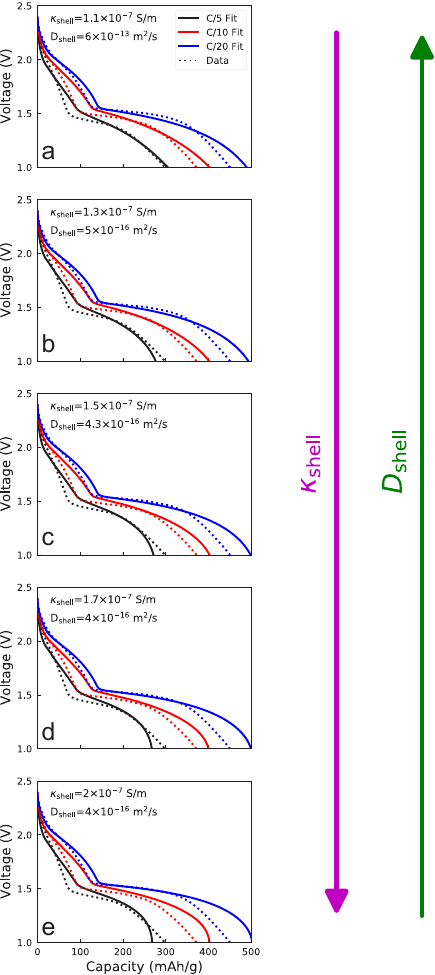}
    {
        \phantomsubcaption\label{fig:ionic:ionic1}
        \phantomsubcaption\label{fig:ionic:ionic2}
        \phantomsubcaption\label{fig:ionic:ionic3}
        \phantomsubcaption\label{fig:ionic:ionic4}
        \phantomsubcaption\label{fig:ionic:ionic5}
    }
    \caption{Parametric comparison for discharge predictions with shell electric conductivity and ionic diffusivity as specified in each figure. Parameters were adjusted while approximately maintaining the C/10 maximum capacity.}
    \label{fig:ionic}
\end{figure}

Our results indicate that electrical transport through the converted shell material is the leading loss mechanism in the conversion regime. However, a major uncertainty in the model surrounds the material properties of the converted shell, as few studies have investigated the composite \ce{Fe}/\ce{Li2S} material and the change in morphology of the converted shell is not taken into account in our model. To study the impact of this uncertainty, we can consider ionic transport through the shell to be a non-negligible factor by decreasing the assumed ionic diffusivity in the shell region. A parametric study on the ionic diffusivity is shown in \cref{fig:ionic}. To maintain agreement with the maximum capacity, we approximately maintain the C/10 maximum capacity in all figures shown in \cref{fig:ionic} by increasing the electrical conductivity to counterbalance the additional ionic transport losses. Our results indicate that with increasing ionic diffusion losses, the discharge curve qualitatively displays a sharper drop-off at the end of the conversion reaction. When compared to the data, non-negligible ionic diffusion losses result in a poorer agreement across different discharge rates which can be seen by comparing the maximum capacities predicted for the C/5 test with the maximum capacities predicted for the C/20 test (\cref{fig:ionic:ionic2,fig:ionic:ionic3,fig:ionic:ionic4,fig:ionic:ionic5}). The poor agreement with experimental data when the shell ionic diffusivity decreases supports our conclusion that electrical transport through the shell is the leading loss mechanism for the conversion regime.

\section{Conclusions}

Throughout this work, we developed a model for conversion cathode materials in a P2D framework which is implemented as an open-source, PyBaMM model. We achieved this by incorporating additional losses through the shell region that can be applied for variables at the reaction surface. Correspondingly, we also incorporated a variable particle radius and specific active surface area to reflect the consumption of small particles as the reaction progresses. The model was able to accurately predict experimental data for \ce{FeS2} cathodes at varying discharge rates and provided insight into the physical processes within the cell. Particularly, we showed that the relative variation in the solid properties across the electrode are small compared to the relative variation in the electrolyte properties across the electrode. However, the magnitude of the electrolyte voltage is small compared to the solid voltage and has a less significant impact on the cell performance. We expect, however, that this electrolyte transport loss would be exacerbated at discharge rates higher than C/5.

The model also proved instrumental to understanding limitations of the cell. We found that at low discharge rates and in the intercalation regime, the leading loss mechanism for the cell was associated with solid-state diffusion polarization, and the leading loss mechanism for the conversion regime was due to shell electrical polarization with some additional loss from kinetic polarization. At faster discharge rates, we found that the intercalation and conversion reactions become mixed, and additional loss mechanisms arise from ionic transport polarization. By incorporating polydispersity into the model and comparing to a monodisperse system, we verified our previous analysis that intercalation materials may effectively be represented as a monodisperse system with radius equal to three times the total volume to total surface area ratio. However, in the conversion regime, increasing polydispersity leads to capacity loss due to the consumption of small particles at the beginning of the reaction.

It should be noted that the model neglects several features that may affect the overall performance. Particularly, we neglect the mechanical deformation of active material particles and the electrode during the conversion reaction, which can result in irreversible particle breakage and cracks. The inability of the current model to account for this expansion may explain some of the deviation with experimental data in the conversion regime. Additionally, the current model fails to account for solid-electrolyte interface (SEI) formation, non-spherical particle geometry, and direct inclusion of the CBD. We intend to explore all these features in a future mesoscopic modeling study. Nevertheless, we believe the model's robustness and efficiency lend itself for use in a variety of other systems such as sodium-ion batteries and other conversion cathode materials (e.g. transition metal fluorides, oxides, selenides, or sulfides). For extension to other conversion chemistry systems, one only needs to modify the appropriate parameters (i.e. open circuit voltage, solid and electrolyte diffusion coefficients, electrical conductivities, equilibrium potential, molar mass, density, and rate constants) and experimental conditions (porosity, electrode thickness, particle radius, electrolyte concentration, and discharge rates). Although we apply the model to \ce{FeS2} in this work, the results presented demonstrate the various insights that can be obtained from the model to better understand cell limitations for other materials.

\section*{Acknowledgments}

We appreciate the interactions and feedback on this work from the entire project team, particularly the leadership and vision of Katharine Harrison and Alec Talin.  We also appreciate internal peer review from Tyler Voskuilen and Harry Moffat prior to submission.

The PyBamm model developed for this paper, along with the associated experimental data, will be made available to the public on Mendeley Data upon publication.  It is available for the reviewers upon request.

A checklist highlighting the completeness of the documentation of this model, following the suggestions of Mistry \textit{et al.} \cite{Mistry2021} can be found in the supplementary information.

This paper describes objective technical results and analysis. Any subjective views or opinions that might be expressed in the paper do not necessarily represent the views of the U.S. Department of Energy or the United States Government. Supported by the Laboratory Directed Research and Development program at Sandia National Laboratories, a multimission laboratory managed and operated by National Technology and Engineering Solutions of Sandia, LLC., a wholly owned subsidiary of Honeywell International, Inc., for the U.S. Department of Energy's National Nuclear Security Administration under contract DE-NA-0003525.

\begin{table}
    \centering
    \caption{Symbol nomenclature.}
    \label{tab:symbol}
    \small
    \begin{tabular}{ll}
        \toprule
        Symbol & Description \\ \midrule
        $a$ & Specific active surface area \\
        $b$ & Bruggeman coefficient \\
        $C$ & Concentration \\
        $C_0$ & Initial electrolyte lithium-ion concentration \\
        $D$ & Diffusion coefficient \\
        $F$ & Faraday constant \\
        $i_\mathrm{cell}$ & Total applied current \\
        $i_0$ & Exchange current density \\
        $j$ & Surface reaction flux \\
        $k$ & Reaction rate constant \\
        $L$ & Length of electrode \\
        $M$ & Molar mass \\
        $r$ & Distance from particle center \\
        $R_c$ & Core radius \\
        $R_p$ & Particle radius \\
        $T$ & Temperature \\
        $t$ & Time \\
        $t_+$ & Transference number \\
        $V$ & Voltage \\
        $x$ & Position across electrode \\
        $z$ & Number of electrons \\
        $\alpha$ & Charge transfer coefficient \\
        $\delta$ & Shell thickness \\
        $\epsilon$ & Electrode porosity \\
        $\nu$ & Overpotential \\
        $\kappa$ & Electrical conductivity \\
        $\mu$ & PSD (\si{\um}) $\ln()$ mean \\
        $\rho$ & Density \\
        $\sigma$ & PSD (\si{\um}) $\ln()$ standard deviation \\
        $\phi$ & Volume fraction \\
        \bottomrule
    \end{tabular}
\end{table}

\begin{table}
    \centering
    \caption{Subscript nomenclature.}
    \label{tab:subscript}
    \small
    \begin{tabular}{ll}
        \toprule
        Subscript & Description \\ \midrule
        CBD & Carbon binder domain property \\
        conv & Conversion reaction \\
        eff & Effective property \\
        eq & Equilibrium \\
        in & Evaluated at reaction surface \\
        int & Intercalation reaction \\
        $l$ & Liquid (electrolyte) property \\
        Li & Intercalated lithium \\
        \ce{Li+} & Lithium-ion \\
        max & Theoretical maximum \\
        out & Evaluated at outer surface of the particle \\
        $s$ & Solid (active material) property \\
        shell & Shell property \\
        \bottomrule
    \end{tabular}
\end{table}

% \bibliographystyle{elsarticle-num} 
% \bibliography{P2D.bib}

\begin{thebibliography}{10}
  \expandafter\ifx\csname url\endcsname\relax
    \def\url#1{\texttt{#1}}\fi
  \expandafter\ifx\csname urlprefix\endcsname\relax\def\urlprefix{URL }\fi
  \expandafter\ifx\csname href\endcsname\relax
    \def\href#1#2{#2} \def\path#1{#1}\fi
  
  \bibitem{Zhang2018}
  X.-Q. Zhang, C.-Z. Zhao, J.-Q. Huang, Q.~Zhang, Recent advances in energy
    chemical engineering of next-generation lithium batteries, Engineering 4~(6)
    (2018) 831--847.
  \newblock \href {http://dx.doi.org/10.1016/j.eng.2018.10.008}
    {\path{doi:10.1016/j.eng.2018.10.008}}.
  
  \bibitem{Wu2020}
  F.~Wu, J.~Maier, Y.~Yu, Guidelines and trends for next-generation rechargeable
    lithium and lithium-ion batteries, Chemical Society Reviews 49~(5) (2020)
    1569--1614.
  \newblock \href {http://dx.doi.org/10.1039/c7cs00863e}
    {\path{doi:10.1039/c7cs00863e}}.
  
  \bibitem{Kim2018}
  J.~Kim, H.~Kim, K.~Kang, Conversion-based cathode materials for rechargeable
    sodium batteries, Advanced Energy Materials 8~(17) (2018) 1702646.
  \newblock \href {http://dx.doi.org/10.1002/aenm.201702646}
    {\path{doi:10.1002/aenm.201702646}}.
  
  \bibitem{Yu2018}
  S.-H. Yu, X.~Feng, N.~Zhang, J.~Seok, H.~D. Abru{\~{n}}a, Understanding
    conversion-type electrodes for lithium rechargeable batteries, Accounts of
    Chemical Research 51~(2) (2018) 273--281.
  \newblock \href {http://dx.doi.org/10.1021/acs.accounts.7b00487}
    {\path{doi:10.1021/acs.accounts.7b00487}}.
  
  \bibitem{Wu2017}
  F.~Wu, G.~Yushin, Conversion cathodes for rechargeable lithium and lithium-ion
    batteries, Energy {\&} Environmental Science 10~(2) (2017) 435--459.
  \newblock \href {http://dx.doi.org/10.1039/c6ee02326f}
    {\path{doi:10.1039/c6ee02326f}}.
  
  \bibitem{Yersak2012}
  T.~A. Yersak, H.~A. Macpherson, S.~C. Kim, V.-D. Le, C.~S. Kang, S.-B. Son,
    Y.-H. Kim, J.~E. Trevey, K.~H. Oh, C.~Stoldt, S.-H. Lee, Solid state enabled
    reversible four electron storage, Advanced Energy Materials 3~(1) (2012)
    120--127.
  \newblock \href {http://dx.doi.org/10.1002/aenm.201200267}
    {\path{doi:10.1002/aenm.201200267}}.
  
  \bibitem{Rickard2007}
  D.~Rickard, G.~W. Luther, Chemistry of iron sulfides, Chemical Reviews 107~(2)
    (2007) 514--562.
  \newblock \href {http://dx.doi.org/10.1021/cr0503658}
    {\path{doi:10.1021/cr0503658}}.
  
  \bibitem{Zou2020}
  J.~Zou, J.~Zhao, B.~Wang, S.~Chen, P.~Chen, Q.~Ran, L.~Li, X.~Wang, J.~Yao,
    H.~Li, J.~Huang, X.~Niu, L.~Wang, Unraveling the reaction mechanism of {FeS}2
    as a li-ion battery cathode, {ACS} Applied Materials {\&} Interfaces 12~(40)
    (2020) 44850--44857.
  \newblock \href {http://dx.doi.org/10.1021/acsami.0c14082}
    {\path{doi:10.1021/acsami.0c14082}}.
  
  \bibitem{Zhang2015}
  S.~S. Zhang, The redox mechanism of {FeS}2 in non-aqueous electrolytes for
    lithium and sodium batteries, Journal of Materials Chemistry A 3~(15) (2015)
    7689--7694.
  \newblock \href {http://dx.doi.org/10.1039/c5ta00623f}
    {\path{doi:10.1039/c5ta00623f}}.
  
  \bibitem{Fong1989}
  R.~Fong, J.~R. Dahn, C.~H.~W. Jones, Electrochemistry of pyrite-based cathodes
    for ambient temperature lithium batteries, Journal of The Electrochemical
    Society 136~(11) (1989) 3206--3210.
  \newblock \href {http://dx.doi.org/10.1149/1.2096426}
    {\path{doi:10.1149/1.2096426}}.
  
  \bibitem{Cabana2010}
  J.~Cabana, L.~Monconduit, D.~Larcher, M.~R. Palac{\'{\i}}n, Beyond
    intercalation-based li-ion batteries: The state of the art and challenges of
    electrode materials reacting through conversion reactions, Advanced Materials
    22~(35) (2010) E170--E192.
  \newblock \href {http://dx.doi.org/10.1002/adma.201000717}
    {\path{doi:10.1002/adma.201000717}}.
  
  \bibitem{Schorr2021}
  B.~N.~B. Schorr, I.~V. Kolesnichenko, L.~C. Merrill, B.~R. Wygant, K.~L.
    Harrison, T.~N. Lambert, Stable cycling of lithium batteries utilizing iron
    disulfide nanoparticles, {ACS} Applied Nano Materials 4~(11) (2021)
    11636--11643.
  \newblock \href {http://dx.doi.org/10.1021/acsanm.1c02178}
    {\path{doi:10.1021/acsanm.1c02178}}.
  
  \bibitem{Cui2022}
  J.~Cui, J.~Liu, X.~Chen, J.~Meng, S.~Wei, T.~Wu, Y.~Wang, Y.~Xie, C.~Lu,
    X.~Zhang, Ganoderma lucidum-derived erythrocyte-like sustainable materials,
    Carbon 196 (2022) 70--77.
  \newblock \href {http://dx.doi.org/10.1016/j.carbon.2022.04.034}
    {\path{doi:10.1016/j.carbon.2022.04.034}}.
  
  \bibitem{Ashby2022}
  D.~S. Ashby, J.~S. Horner, G.~Whang, A.~S. Lapp, S.~A. Roberts, B.~Dunn, I.~V.
    Kolesnichenko, T.~N. Lambert, A.~A. Talin, Understanding the electrochemical
    performance of {FeS$_2$} conversion cathodes, {ACS} Applied Materials {\&}
    Interfaces 14~(23) (2022) 26604--26611.
  \newblock \href {http://dx.doi.org/10.1021/acsami.2c01021}
    {\path{doi:10.1021/acsami.2c01021}}.
  
  \bibitem{Su2018}
  Q.~Su, Y.~Lu, S.~Liu, X.~Zhang, Y.~Lin, R.~Fu, D.~Wu, Nanonetwork-structured
    yolk-shell {FeS}2\@c as high-performance cathode materials for li-ion
    batteries, Carbon 140 (2018) 433--440.
  \newblock \href {http://dx.doi.org/10.1016/j.carbon.2018.08.049}
    {\path{doi:10.1016/j.carbon.2018.08.049}}.
  
  \bibitem{Roberts2016}
  S.~A. Roberts, H.~Mendoza, V.~E. Brunini, B.~L. Trembacki, D.~R. Noble, A.~M.
    Grillet, Insights into lithium-ion battery degradation and safety mechanisms
    from mesoscale simulations using experimentally reconstructed mesostructures,
    Journal of Electrochemical Energy Conversion and Storage 13~(3).
  \newblock \href {http://dx.doi.org/10.1115/1.4034410}
    {\path{doi:10.1115/1.4034410}}.
  
  \bibitem{Zhang2000}
  D.~Zhang, B.~N. Popov, R.~E. White, Modeling lithium intercalation of a single
    spinel particle under potentiodynamic control, Journal of The Electrochemical
    Society 147~(3) (2000) 831.
  \newblock \href {http://dx.doi.org/10.1149/1.1393279}
    {\path{doi:10.1149/1.1393279}}.
  
  \bibitem{Doyle1993}
  M.~Doyle, T.~F. Fuller, J.~Newman, Modeling of galvanostatic charge and
    discharge of the lithium/polymer/insertion cell, Journal of The
    Electrochemical Society 140~(6) (1993) 1526--1533.
  \newblock \href {http://dx.doi.org/10.1149/1.2221597}
    {\path{doi:10.1149/1.2221597}}.
  
  \bibitem{Santhanagopalan2006}
  S.~Santhanagopalan, Q.~Guo, P.~Ramadass, R.~E. White, Review of models for
    predicting the cycling performance of lithium ion batteries, Journal of Power
    Sources 156~(2) (2006) 620--628.
  \newblock \href {http://dx.doi.org/10.1016/j.jpowsour.2005.05.070}
    {\path{doi:10.1016/j.jpowsour.2005.05.070}}.
  
  \bibitem{Ramadesigan2012}
  V.~Ramadesigan, P.~W.~C. Northrop, S.~De, S.~Santhanagopalan, R.~D. Braatz,
    V.~R. Subramanian, Modeling and simulation of lithium-ion batteries from a
    systems engineering perspective, Journal of The Electrochemical Society
    159~(3) (2012) R31--R45.
  \newblock \href {http://dx.doi.org/10.1149/2.018203jes}
    {\path{doi:10.1149/2.018203jes}}.
  
  \bibitem{Marquis2019}
  S.~G. Marquis, V.~Sulzer, R.~Timms, C.~P. Please, S.~J. Chapman, An asymptotic
    derivation of a single particle model with electrolyte, Journal of The
    Electrochemical Society 166~(15) (2019) A3693--A3706.
  \newblock \href {http://dx.doi.org/10.1149/2.0341915jes}
    {\path{doi:10.1149/2.0341915jes}}.
  
  \bibitem{Srinivasan2004}
  V.~Srinivasan, J.~Newman, Discharge model for the lithium iron-phosphate
    electrode, Journal of The Electrochemical Society 151~(10) (2004) A1517.
  \newblock \href {http://dx.doi.org/10.1149/1.1785012}
    {\path{doi:10.1149/1.1785012}}.
  
  \bibitem{Subramanian2000}
  V.~R. Subramanian, H.~J. Ploehn, R.~E. White, Shrinking core model for the
    discharge of a metal hydride electrode, Journal of The Electrochemical
    Society 147~(8) (2000) 2868.
  \newblock \href {http://dx.doi.org/10.1149/1.1393618}
    {\path{doi:10.1149/1.1393618}}.
  
  \bibitem{Jagannathan2009}
  K.~Jagannathan, Approximate solution methods for solid-state diffusion in
    phase-change electrodes, Journal of The Electrochemical Society 156~(12)
    (2009) A1028.
  \newblock \href {http://dx.doi.org/10.1149/1.3237142}
    {\path{doi:10.1149/1.3237142}}.
  
  \bibitem{Zhu2010}
  Y.~Zhu, C.~Wang, Novel {CV} for phase transformation electrodes, The Journal of
    Physical Chemistry C 115~(3) (2010) 823--832.
  \newblock \href {http://dx.doi.org/10.1021/jp109954y}
    {\path{doi:10.1021/jp109954y}}.
  
  \bibitem{Prada2012}
  E.~Prada, D.~D. Domenico, Y.~Creff, J.~Bernard, V.~Sauvant-Moynot, F.~Huet,
    Simplified electrochemical and thermal model of {LiFePO}4-graphite li-ion
    batteries for fast charge applications, Journal of The Electrochemical
    Society 159~(9) (2012) A1508--A1519.
  \newblock \href {http://dx.doi.org/10.1149/2.064209jes}
    {\path{doi:10.1149/2.064209jes}}.
  
  \bibitem{Horner2021}
  J.~S. Horner, G.~Whang, D.~S. Ashby, I.~V. Kolesnichenko, T.~N. Lambert, B.~S.
    Dunn, A.~A. Talin, S.~A. Roberts, Electrochemical modeling of {GITT}
    measurements for improved solid-state diffusion coefficient evaluation, {ACS}
    Applied Energy Materials 4~(10) (2021) 11460--11469.
  \newblock \href {http://dx.doi.org/10.1021/acsaem.1c02218}
    {\path{doi:10.1021/acsaem.1c02218}}.
  
  \bibitem{Vijayasekaran2006}
  B.~Vijayasekaran, C.~A. Basha, Shrinking core discharge model for the negative
    electrode of a lead-acid battery, Journal of Power Sources 158~(1) (2006)
    710--721.
  \newblock \href {http://dx.doi.org/10.1016/j.jpowsour.2005.10.006}
    {\path{doi:10.1016/j.jpowsour.2005.10.006}}.
  
  \bibitem{Bruggeman1935}
  D.~A.~G. Bruggeman, Berechnung verschiedener physikalischer konstanten von
    heterogenen substanzen. i. dielektrizitätskonstanten und leitfähigkeiten
    der mischkörper aus isotropen substanzen, Annalen der Physik 416~(7) (1935)
    636--664.
  \newblock \href {http://dx.doi.org/10.1002/andp.19354160705}
    {\path{doi:10.1002/andp.19354160705}}.
  
  \bibitem{Andersson2018}
  J.~A.~E. Andersson, J.~Gillis, G.~Horn, J.~B. Rawlings, M.~Diehl, {CasADi}: a
    software framework for nonlinear optimization and optimal control,
    Mathematical Programming Computation 11~(1) (2018) 1--36.
  \newblock \href {http://dx.doi.org/10.1007/s12532-018-0139-4}
    {\path{doi:10.1007/s12532-018-0139-4}}.
  
  \bibitem{Harris2020}
  C.~R. Harris, K.~J. Millman, S.~J. van~der Walt, R.~Gommers, P.~Virtanen,
    D.~Cournapeau, E.~Wieser, J.~Taylor, S.~Berg, N.~J. Smith, R.~Kern, M.~Picus,
    S.~Hoyer, M.~H. van Kerkwijk, M.~Brett, A.~Haldane, J.~F. del R{\'{\i}}o,
    M.~Wiebe, P.~Peterson, P.~G{\'{e}}rard-Marchant, K.~Sheppard, T.~Reddy,
    W.~Weckesser, H.~Abbasi, C.~Gohlke, T.~E. Oliphant, Array programming with
    {NumPy}, Nature 585~(7825) (2020) 357--362.
  \newblock \href {http://dx.doi.org/10.1038/s41586-020-2649-2}
    {\path{doi:10.1038/s41586-020-2649-2}}.
  
  \bibitem{Sulzer2020}
  V.~Sulzer, S.~G. Marquis, R.~Timms, M.~Robinson, S.~J. Chapman, Python battery
    mathematical modelling ({PyBaMM})\href
    {http://dx.doi.org/10.1149/osf.io/67ckj} {\path{doi:10.1149/osf.io/67ckj}}.
  
  \bibitem{Castiglione2011}
  F.~Castiglione, E.~Ragg, A.~Mele, G.~B. Appetecchi, M.~Montanino, S.~Passerini,
    Molecular environment and enhanced diffusivity of {Li+} ions in
    lithium-salt-doped ionic liquid electrolytes, The Journal of Physical
    Chemistry Letters 2~(3) (2011) 153--157.
  \newblock \href {http://dx.doi.org/10.1021/jz101516c}
    {\path{doi:10.1021/jz101516c}}.
  
  \bibitem{Pyrite}
  M.~Minerals, \href{http://www.webmineral.com/data/Pyrite.shtml}{Pyrite mineral
    data}.
  \newline\urlprefix\url{http://www.webmineral.com/data/Pyrite.shtml}
  
  \bibitem{Yan2021}
  S.~Yan, Y.~Wang, T.~Chen, Z.~Gan, S.~Chen, Y.~Liu, S.~Zhang, Regulated
    interfacial stability by coordinating ionic liquids with fluorinated solvent
    for high voltage and safety batteries, Journal of Power Sources 491 (2021)
    229603.
  \newblock \href {http://dx.doi.org/10.1016/j.jpowsour.2021.229603}
    {\path{doi:10.1016/j.jpowsour.2021.229603}}.
  
  \bibitem{Nadherna2011}
  M.~N{\'{a}}dhern{\'{a}}, J.~Reiter, J.~Mo{\v{s}}kon, R.~Dominko, Lithium
    bis(fluorosulfonyl)imide{\textendash}{PYR}14tfsi ionic liquid electrolyte
    compatible with graphite, Journal of Power Sources 196~(18) (2011)
    7700--7706.
  \newblock \href {http://dx.doi.org/10.1016/j.jpowsour.2011.04.033}
    {\path{doi:10.1016/j.jpowsour.2011.04.033}}.
  
  \bibitem{Son2013}
  S.-B. Son, T.~A. Yersak, D.~M. Piper, S.~C. Kim, C.~S. Kang, J.~S. Cho, S.-S.
    Suh, Y.-U. Kim, K.~H. Oh, S.-H. Lee, A stabilized {PAN}-{FeS}2cathode with an
    {EC}/{DEC} liquid electrolyte, Advanced Energy Materials 4~(3) (2013)
    1300961.
  \newblock \href {http://dx.doi.org/10.1002/aenm.201300961}
    {\path{doi:10.1002/aenm.201300961}}.
  
  \bibitem{Park2013}
  J.-W. Park, K.~Ueno, N.~Tachikawa, K.~Dokko, M.~Watanabe, Ionic liquid
    electrolytes for lithium{\textendash}sulfur batteries, The Journal of
    Physical Chemistry C 117~(40) (2013) 20531--20541.
  \newblock \href {http://dx.doi.org/10.1021/jp408037e}
    {\path{doi:10.1021/jp408037e}}.
  
  \bibitem{Evans2014}
  T.~Evans, D.~M. Piper, S.~C. Kim, S.~S. Han, V.~Bhat, K.~H. Oh, S.-H. Lee,
    Ionic liquid enabled {FeS}2for high-energy-density lithium-ion batteries,
    Advanced Materials 26~(43) (2014) 7386--7392.
  \newblock \href {http://dx.doi.org/10.1002/adma.201402103}
    {\path{doi:10.1002/adma.201402103}}.
  
  \bibitem{Mistry2021}
  A.~Mistry, A.~Verma, S.~Sripad, R.~Ciez, V.~Sulzer, F.~B. Planella, R.~Timms,
    Y.~Zhang, R.~Kurchin, P.~Dechent, W.~Li, S.~Greenbank, Z.~Ahmad,
    D.~Krishnamurthy, A.~M. Fenton, K.~Tenny, P.~Patel, D.~J. Robles, P.~Gasper,
    A.~Colclasure, A.~Baskin, C.~D. Scown, V.~R. Subramanian, E.~Khoo, S.~Allu,
    D.~Howey, S.~DeCaluwe, S.~A. Roberts, V.~Viswanathan, A minimal information
    set to enable verifiable theoretical battery research, {ACS} Energy Letters
    6~(11) (2021) 3831--3835.
  \newblock \href {http://dx.doi.org/10.1021/acsenergylett.1c01710}
    {\path{doi:10.1021/acsenergylett.1c01710}}.
  
  \end{thebibliography}

\end{document}

\begin{frontmatter}

\title{Supplementary Information: A Pseudo-Two-Dimensional (P2D) Model for \ce{FeS2} Conversion Cathode Batteries}

\author[1]{Jeffrey S. Horner}
\author[2]{Grace Whang}
\author[3]{Igor V. Kolesnichenko}
\author[3]{Timothy N. Lambert}
\author[2]{Bruce S. Dunn}
\author[1]{Scott A. Roberts\corref{cor1}}
\ead{sarober@sandia.gov}
\cortext[cor1]{Corresponding author}
\address[1]{Thermal/Fluid Component Sciences Department, Sandia National Laboratories, Albuquerque, New Mexico, USA}
\address[2]{Materials Science and Engineering Department, University of California, Los Angeles, Los Angeles, California, USA}
\address[3]{Photovoltaics and Materials Technology Department, Sandia National Laboratories, Albuquerque, New Mexico, USA}

\begin{keyword}

Conversion cathode materials \sep lithium-ion battery \sep \ce{FeS2} \sep pseudo-two-dimensional (P2D) modeling \sep electrical conductivity \sep ionic diffusivity

\end{keyword}

\end{frontmatter}

\section{GITT Measurements}
 
\begin{figure*}
    \centering
    \begin{subfigure}{0.67\linewidth}
        \includegraphics[width=\linewidth]{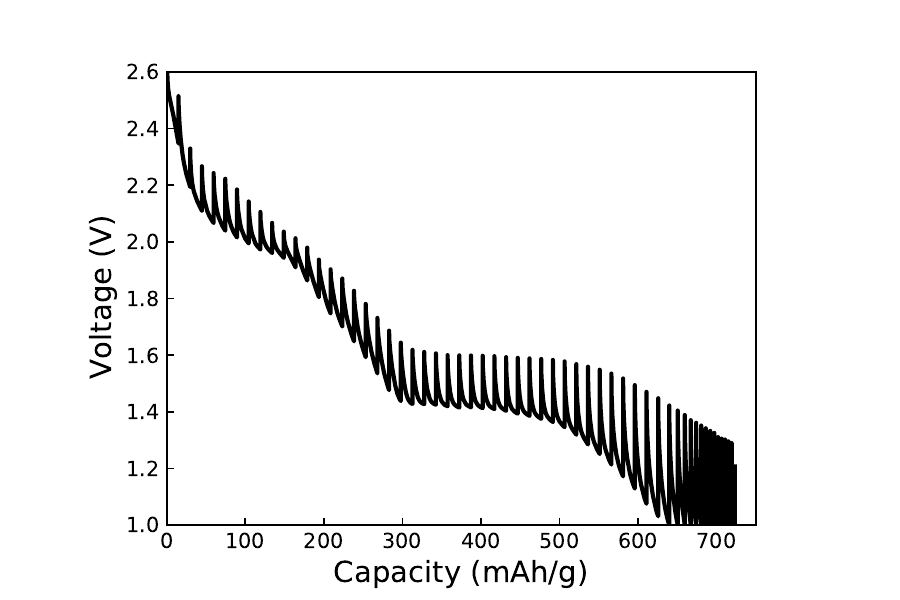}
        \phantomcaption\label{fig:methods:OCV}
    \end{subfigure}
    \caption{\textbf{GITT measurements on \ce{FeS2} from 2.4-1.0 V.} For each pulse, the cell was subjected to C/20 for 20 minutes followed by a 4 hour rest step.}
    \label{fig:gitt}
\end{figure*}

 To evaluate the open circuit voltage (OCV) and reaction rate constants $k$ in the model, we performed galvanostatic intermittent titration technique (GITT) measurements on the cell from 2.4-\SI{1.0}{V} according to the protocol described in \cite{Horner2021}. Note, the 2.4-\SI{1.0}{V} measurements were only used to evaluate the conversion reaction properties, and the intercalation reaction properties were all evaluated from the previously published GITT measurements on the isolated intercalation regime \cite{Horner2021}. The experimental results are shown in \cref{fig:gitt}. We evaluated the OCV by exponentially extrapolating the rest steps:
 \begin{equation}
    \ln\left(OCV-V\right) = k_1-\frac{t}{\tau},
    \label{eq:OCV}
\end{equation}
where $V$ is the voltage, $k_1$ and $\tau$ are constants, and $t$ is the step time for the relaxation. At \SI{1.6}{V}, we observed a plateau in the OCV indicating the conversion reaction has begun. Thus, we elected to use a constant value of \SI{1.6}{V} for the conversion OCV. Note that the decrease in OCV at the end is likely due to deviation from exponential behavior arising from additional sources of transport limitations. For the intercalation regime, we fit a high order polynomial to the OCV data:
\begin{equation}
    \begin{aligned}
    OCV&=-56.192\left(\frac{2C_\mathrm{Li}}{C_\mathrm{max}}\right)^6+506.297\left(\frac{2C_\mathrm{Li}}{C_\mathrm{max}}\right)^5-1882.986\left(\frac{2C_\mathrm{Li}}{C_\mathrm{max}}\right)^4 \\ &+3692.354\left(\frac{2C_\mathrm{Li}}{C_\mathrm{max}}\right)^3-4016.451\left(\frac{2C_\mathrm{Li}}{C_\mathrm{max}}\right)^2+2290.362\left(\frac{2C_\mathrm{Li}}{C_\mathrm{max}}\right) \\ 
    &-530.220,
    \end{aligned}
\end{equation}
where the OCV is evaluated in units of volts.

We evaluated the rate constant for each GITT pulse from the initial change in voltage (over the first 2 seconds) upon switching the current on/off by first solving:
\begin{equation}
    \Delta i_\mathrm{cell}=2i_{0,i}a\sinh{\left(\frac{\alpha F\Delta V}{RT}\right)},
\end{equation}
for the exchange current density $i_{0,i}$. This value can then be used to determine the rate constants for the intercalation and conversion reactions by using their respective definitions for the exchange current densities provided in the main text. As we determined separate rate constants for each GITT pulse, the single value used throughout the main text is a median of all obtained values. Note that we only used GITT pulses entirely in the intercalation regime (2.4-\SI{1.6}{V}) and entirely in the conversion regime (1.6-\SI{1.0}{V}) for the respective rate constants.

\section{Conversion Reaction Charge Transfer Coefficient}

\begin{figure*}
    \centering
    \begin{subfigure}{0.67\linewidth}
        \includegraphics[width=\linewidth]{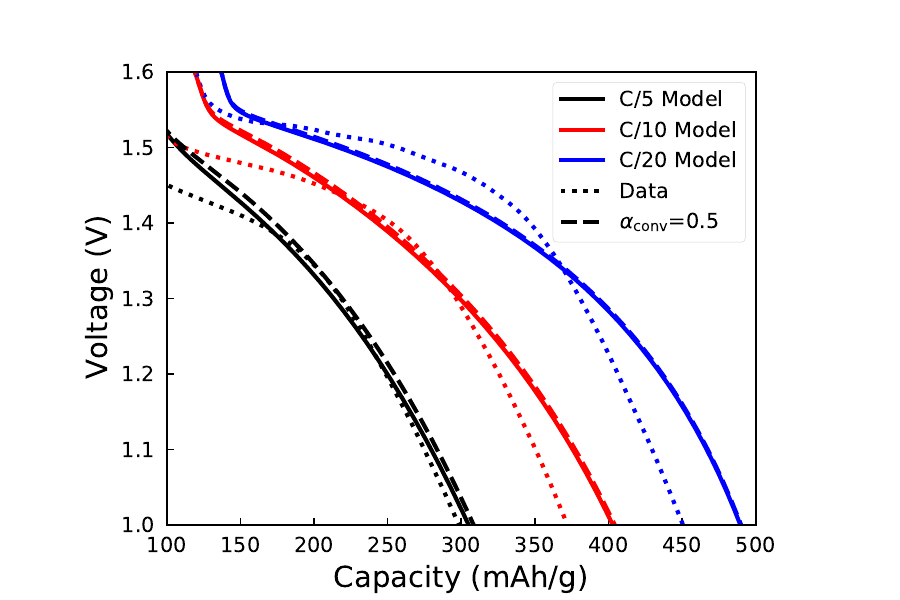}
        \phantomcaption\label{fig:alpha:crates_alpha}
    \end{subfigure}
    \caption{\textbf{Conversion reaction dependence on the charge transfer coefficient.} Solid lines represent a value of 2 for the conversion charge transfer coefficient ($\alpha_\mathrm{conv}$), dashed lines represent a value of 0.5, and the dotted lines represent the experimental data.}
    \label{fig:alpha}
\end{figure*}

The exchange current density for the conversion reaction is defined as:
\begin{equation}
    i_{0,\mathrm{conv}}=k_\mathrm{conv}F\left(C_{\ce{Li+},\mathrm{in}}\right)^{\alpha_\mathrm{conv}}
\end{equation}
where $\alpha_\mathrm{conv}$ corresponds to the charge transfer coefficient for the conversion reaction. Throughout the main text, we use a value of 2 for this parameter. A comparison between a value of 2 (indicated as the solid lines) and a value of 0.5 (indicated as the dashed lines) is shown in \cref{fig:alpha} compared to the experimental data. Overall, this parameter plays a small role in the model behavior. However, a value of 2 gives slightly better results than that of 0.5 which was used for the intercalation reaction.

\section{Shell Pseudo-Steady-State Transport Derivations}

To approximate the transport losses through the converted shell, we used pseudo-steady-state approximations for the change in solid voltage, electrolyte voltage, and lithium-ion concentration. The solid voltage change is derived from Ohm's law for current conservation in spherical coordinates:
\begin{equation}
    \frac{1}{r^2}\frac{\partial}{\partial r}\left(r^2\kappa_\mathrm{shell} \frac{\partial V_s}{\partial r}\right)=0,
\end{equation}
with the boundary conditions:
\begin{equation}
    V_s\left(r=R_p\right)=V_{s,\mathrm{out}}
\end{equation}
\begin{equation}
    \kappa_\mathrm{shell}\frac{\partial V_s}{\partial r}\Bigr|_{\substack{r=R_c}} = jF.
\end{equation}
Upon solving the above differential equation, we obtain the solution for the solid voltage throughout the shell:
\begin{equation}
    V_s(r)=V_{s,\mathrm{out}}-\frac{jFR_c^2}{\kappa_\mathrm{shell}}\left(\frac{1}{r}-\frac{1}{R_p}\right),
\end{equation}
which can be applied at the reaction surface $R_c$ to obtain the voltage drop.

The electrolyte voltage drop and change in lithium-ion concentration are derived from the differential, steady-state, Nernst-Planck equation in spherical coordinates:
\begin{equation}
    \frac{1}{r^2}\frac{\partial}{\partial r}\left[r^2D_{i,\mathrm{shell}}\left(\frac{\partial C_i}{\partial r}+\frac{z_iFC_i}{RT}\frac{\partial V_l}{\partial r}\right)\right]=0,
\end{equation}
where $i$ indicates the ionic species. If we assume a two-ion system with electroneutrality for simplicity, we achieve the necessary condition $C_\mathrm{Li^+}=C_\mathrm{FSI^-}$ to solve the above equation for both species subject to the boundary conditions:
\begin{equation}
    V_l\left(r=R_p\right)=V_{l,\mathrm{out}}
\end{equation}
\begin{equation}
    C_\mathrm{Li^+}\left(r=R_p\right)=C_\mathrm{Li^+,out}
\end{equation}
\begin{equation}
    D_\mathrm{Li^+,shell}\left(\frac{\partial C_\mathrm{Li^+}}{\partial r}+\frac{FC_\mathrm{Li^+}}{RT}\frac{\partial V_l}{\partial r}\right)\Bigr|_{\substack{r=R_c}} =-j
\end{equation}
\begin{equation}
    D_\mathrm{FSI^-,shell}\left(\frac{\partial C_\mathrm{FSI^-}}{\partial r}+\frac{FC_\mathrm{Li^+}}{RT}\frac{\partial V_l}{\partial r}\right)\Bigr|_{\substack{r=R_c}} =0.
\end{equation}
Solving this system for the ionic lithium concentration and electrolyte voltage throughout the shell gives:
\begin{equation}
    C_\mathrm{Li^+}(r)=C_\mathrm{\ce{Li+},out}+\frac{jR_c^2}{2D_\mathrm{Li^+,shell}}\left(\frac{1}{r}-\frac{1}{R_p}\right) \label{eq:concloss}
\end{equation}
\begin{equation}
    V_l(r)=V_{l,\mathrm{out}}+\ln\left(\frac{C_\mathrm{Li^+}(r)}{C_\mathrm{\ce{Li+},out}}\right)\frac{RT}{F} \label{eq:elecloss},
\end{equation}
which will reduce to the equations provided in the main text when applied at the reaction surface $R_c$. Note that the two-ion assumption differs from the assumptions of the DFN model used throughout the electrode. Nevertheless, we do not anticipate that this alternative assumption significantly affects the results as the voltage and ionic concentration drop throughout the shell were found to have little impact on the overall results.

\section{Polydisperse Derivations}

We used a log-normal distribution to represent our particle size distribution, where the probability density function $f$ for particle radii $R_p$ is given by:
\begin{equation}
    f(R_p)=\frac{1}{R_p\sigma\sqrt{2\pi}}\exp{\left[-\frac{\left(\ln{\left(R_p\right)}-\mu\right)^2}{2\sigma^2}\right]}.
\end{equation}
Here, $\sigma$ is the PSD (\si{\um}) $\ln()$ standard deviation, and $\mu$ is the PSD (\si{\um}) $\ln()$ mean. During conversion, the effective particle radius $R_{p,\mathrm{eff}}$ can be evaluated from the distribution as:
\begin{equation}
    R_{p,\mathrm{eff}}=\frac{3V_\mathrm{Tot,out}(\delta)}{A_\mathrm{Tot,out}(\delta)}=\frac{\int_{\delta}^{\infty} f(R_p)R_p^3dR_p}{\int_{\delta}^{\infty} f(R_p)R_p^2dR_p},
    \label{eq:Rpeff}
\end{equation}
where $\delta$ is the thickness of the converted shell. Note the lower bound of the integrals represents the maximum particle radius that is still active in the reaction as smaller particles will be entirely consumed prior to larger particles. It is also important to note that the effective particle radius defined here corresponds to the the effective outer radius for all particles still active, not the inner core radius. The shrinking of the core is captured in the specific active surface area $a$ which can be evaluated as:
\begin{equation}
    a=(1-\epsilon)\frac{A_\mathrm{Tot,in}(\delta)}{V_\mathrm{Tot,out}(\delta=0)}=(1-\epsilon)\frac{3\int_{\delta}^{\infty} f(R_p)\left(R_p-\delta\right)^2dR_p}{\int_{0}^{\infty} f(R_p)R_p^3dR_p}.
    \label{eq:a}
\end{equation}

All relevant integrals can be evaluated analytically for the log-normal distribution as:
\begin{equation}
    \int_{\delta}^{\infty} f(R_p)R_p^3dR_p=\frac{1}{2}\exp{\left(3\mu+\frac{9\sigma^2}{2}\right)}\left[1+\mathrm{erf}\left(\frac{\mu+3\sigma^2-\ln{\left(\delta\right)}}{\sqrt{2}\sigma}\right)\right]
\end{equation}
\begin{equation}
    \int_{\delta}^{\infty} f(R_p)R_p^2dR_p=\frac{1}{2}\exp{\left(2\mu+2\sigma^2\right)}\left[1+\mathrm{erf}\left(\frac{\mu+2\sigma^2-\ln{\left(\delta\right)}}{\sqrt{2}\sigma}\right)\right]
\end{equation}
\begin{equation}
    \begin{aligned}
    \int_{\delta}^{\infty} f(R_p)\left(R_p-\delta\right)^2dR_p&=\frac{\delta^2}{2}\left[1+\mathrm{erf}\left(\frac{\mu-\ln{\left(\delta\right)}}{\sqrt{2}\sigma}\right)\right] \\ &-\exp{\left(\mu+\frac{\sigma^2}{2}\right)}\left[1+\mathrm{erf}\left(\frac{\mu+\sigma^2-\ln{\left(\delta\right)}}{\sqrt{2}\sigma}\right)\right] \\ &+\frac{1}{2}\exp{\left(2\mu+2\sigma^2\right)}\left[1+\mathrm{erf}\left(\frac{\mu+2\sigma^2-\ln{\left(\delta\right)}}{\sqrt{2}\sigma}\right)\right]
    \end{aligned}
\end{equation}
\begin{equation}
    \int_{0}^{\infty} f(R_p)R_p^3dR_p=\exp{\left(3\mu+\frac{9\sigma^2}{2}\right)}.
\end{equation}
By substituting the above integrals into \cref{eq:Rpeff} and \cref{eq:a}, we can obtain analytical expressions for the effective particle radius and specific active surface area, respectively, as a function of the converted shell thickness. The expressions can be recast as a function of the extent of conversion reaction $\xi_\mathrm{conv}$ by numerically solving for $\delta$ as a function of $\xi_\mathrm{conv}$ according to:
\begin{equation}
    V_\mathrm{Tot,out}(\xi_\mathrm{conv}=0)=4\pi\int_{\delta(\xi_\mathrm{conv})}^{\infty} f(R_p)\left(R_p-\delta(\xi_\mathrm{conv})\right)^2\frac{d\delta}{d\xi_\mathrm{conv}}dR_p.
\end{equation}

\section{Model Checklist}

\begin{table}[H]
    \centering
    \caption{Model checklist from \cite{Mistry2021}.}
    \label{tab:checklist}
    \small
    \begin{tabular}{p{0.85\linewidth} >{\centering\arraybackslash}p{0.1\linewidth}}
        \toprule
        \multicolumn{2}{p{0.95\linewidth}}{\textbf{Manuscript Title:} A Pseudo-Two-Dimensional (P2D) Conversion Chemistry Model for \ce{FeS2} Cathode Lithium-Ion Batteries} \\ \midrule
        \textbf{Submitting Author*:} Scott A. Roberts \\ \midrule
        \textbf{Question:} & \textbf{Y/N/NA} \\ \midrule
        
        1. Have you provided all assumptions, theory, governing equations, initial and boundary conditions, material properties (e.g., open-circuit potential) with appropriate precision and literature sources, constant states (e.g., temperature), etc.? & Y \\
        \multicolumn{2}{p{0.95\linewidth}}{\textbf{Remarks:}} \\ \midrule
        
        2. If the calculations have a probabilistic component (e.g., Monte Carlo, initial configuration in Molecular Dynamics, etc.), did you provide statistics (mean, standard deviation, confidence interval, etc.) from multiple (≥3) runs of a representative case? & NA \\
        \multicolumn{2}{p{0.95\linewidth}}{\textbf{Remarks:} There is no probabilistic component to this work, and we do not consider uncertainties or variabilities of any model parameters.} \\ \midrule
        
        3. If data-driven calculations are performed (e.g., machine learning), did you specify dataset origin, the rationale behind choosing it, what information it contains, and the specific portion of it being utilized? Have you described the thought process for choosing a specific modeling paradigm? & NA \\
        \multicolumn{2}{p{0.95\linewidth}}{\textbf{Remarks:} No data-driven modeling is included.  Comparisons to experimental data is thoroughly discussed. } \\ \midrule
        
        4. Have you discussed all sources of potential uncertainty, variability, and errors in the modeling results and their impact on quantitative results and qualitative trends? Have you discussed the sensitivity of modeling (and numerical) inputs such as material properties, time step, domain size, neural network architecture, etc. where they are variable or uncertain? & N \\
        \multicolumn{2}{p{0.95\linewidth}}{\textbf{Remarks:} Time step and domain size were found to have no effect on the results. There are surely uncertainties in many of the model parameters, but a careful study of that is outside of the scope of this paper.} \\ \midrule
        
        5. Have you sufficiently discussed new or not widely familiar terminology and descriptors for clarity? Did you use these terms in their appropriate context to avoid misinterpretation? Enumerate these terms in the “Remarks”. & Y \\
        \multicolumn{2}{p{0.95\linewidth}}{\textbf{Remarks:} The only not widely familiar terminology may be the polarization losses outlined in Table 4 of the main text.} \\ \midrule
        
        \multicolumn{2}{p{0.95\linewidth}}{*I verify that this form is completed accurately in agreement with all co-authors, to the best of my knowledge. a Y $=$ the question is answered completely. Discuss any N or NA response in “Remarks”.} \\ \bottomrule
    \end{tabular}
\end{table}

\section{C/10 and C/20 Results}

\begin{figure*}
    \centering
    {
        \phantomsubcaption\label{fig:C10:C10}
        \phantomsubcaption\label{fig:C10:contourC10}
        \phantomsubcaption\label{fig:C10:conversionC10}
        \phantomsubcaption\label{fig:C10:shellC10}
        \phantomsubcaption\label{fig:C10:lithiumC10}
        \phantomsubcaption\label{fig:C10:solidC10}
        \phantomsubcaption\label{fig:C10:lithium-ionC10}
        \phantomsubcaption\label{fig:C10:electrolyteC10}
    }
    \includegraphics[width=\linewidth]{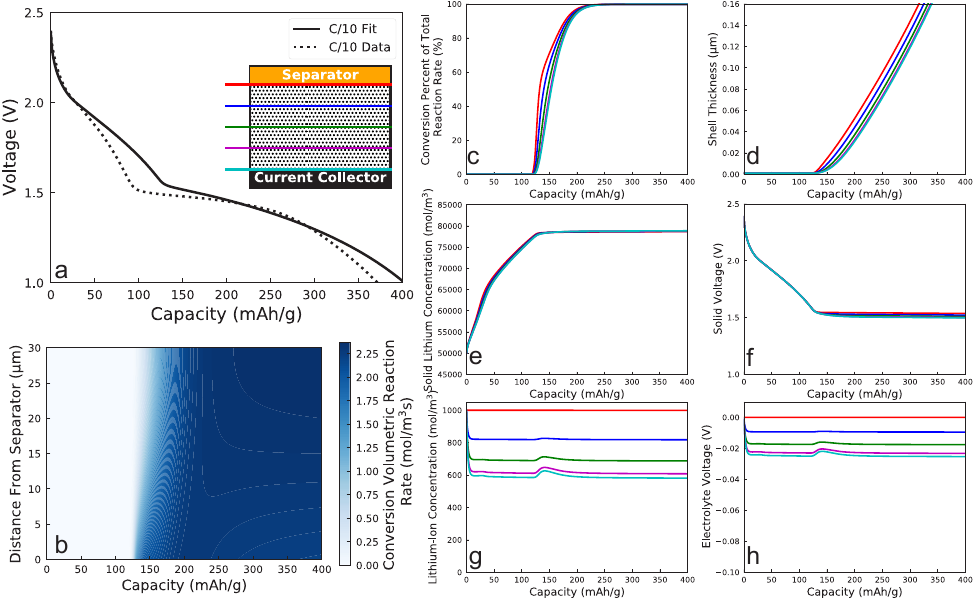}
    \caption{\textbf{Model predictions and comparison to experimental data for a C/10 discharge rate.} (a) C/10 experimental discharge data and model prediction. Inset shows a schematic of the cell with colored lines used in (c)-(h) marking their different locations across the cell. The texture within the cell reflects the porous electrode/electrolyte system with a thickness of \SI{30}{\micro m}. (b) Contour map showing the volumetric reaction rate for the conversion reaction as a function of the specific capacity and distance from the separator. (c) Percent of total reaction rate that corresponds to the conversion reaction, (d) shell thickness, (e) solid lithium concentration, (f) solid voltage, (g) lithium-ion concentration in the electrolyte phase, and (h) electrolyte voltage are all shown as a function of capacity and evaluated at the reaction surface.}
    \label{fig:C10}
\end{figure*}

The model comparison with experimental data and predictions for various quantities of interest at different distances across the electrode as a function of capacity are shown in \cref{fig:C10} and \cref{fig:C10} for C/10 and C/20, respectively. The results for the slower discharge rates agree with those presented in the main text for C/5 in that the most significant changes across the electrode manifest in the shell thickness and lithium-ion concentration. However, at slower discharge rates, these variations across the electrode become less significant and the cell as a whole becomes less polarized. The reduction of polarization can particularly be seen in the solid voltage and conversion percent of total reaction rate, both of which are nearly uniform across the electrode at all times for a C/20 discharge rate in contrast to the results shown in the main text for C/5.

\begin{figure*}
    \centering
    {
        \phantomsubcaption\label{fig:C20:C20}
        \phantomsubcaption\label{fig:C20:contourC20}
        \phantomsubcaption\label{fig:C20:conversionC20}
        \phantomsubcaption\label{fig:C20:shellC20}
        \phantomsubcaption\label{fig:C20:lithiumC20}
        \phantomsubcaption\label{fig:C20:solidC20}
        \phantomsubcaption\label{fig:C20:lithium-ionC20}
        \phantomsubcaption\label{fig:C20:electrolyteC20}
    }
    \includegraphics[width=\linewidth]{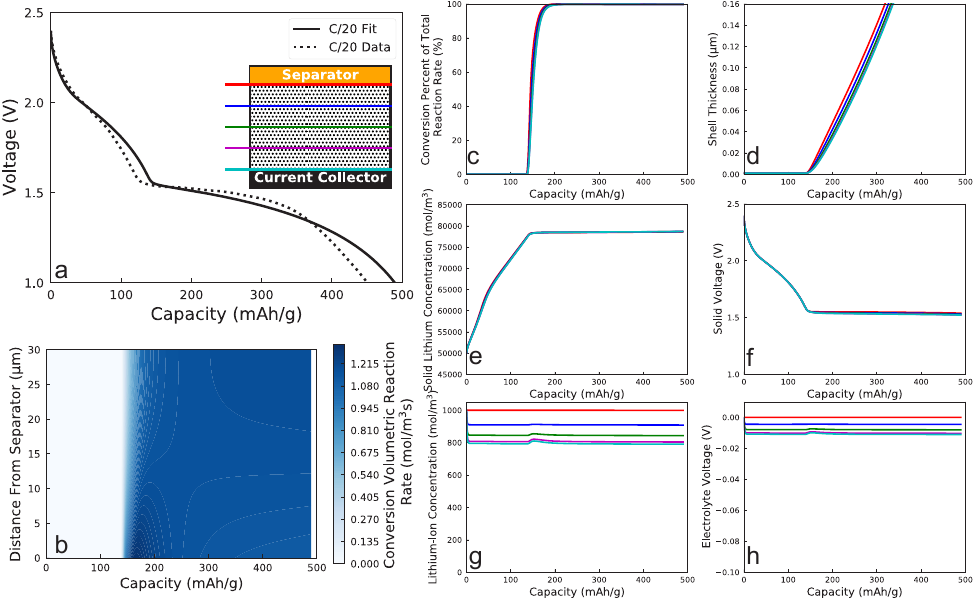}
    \caption{\textbf{Model predictions and comparison to experimental data for a C/20 discharge rate.} (a) C/20 experimental discharge data and model prediction. Inset shows a schematic of the cell with colored lines used in (c)-(h) marking their different locations across the cell. The texture within the cell reflects the porous electrode/electrolyte system with a thickness of \SI{30}{\micro m}. (b) Contour map showing the volumetric reaction rate for the conversion reaction as a function of the specific capacity and distance from the separator. (c) Percent of total reaction rate that corresponds to the conversion reaction, (d) shell thickness, (e) solid lithium concentration, (f) solid voltage, (g) lithium-ion concentration in the electrolyte phase, and (h) electrolyte voltage are all shown as a function of capacity and evaluated at the reaction surface.}
    \label{fig:C20}
\end{figure*}
 
% \bibliographystyle{elsarticle-num}
% \bibliography{P2D.bib}